\documentclass[useAMS,usenatbib,subeqn]{mn2e}

\usepackage{multirow}  
\usepackage{color}
\usepackage{epsfig,amssymb}
\usepackage{graphicx,epsfig}
 \usepackage{hhline}
\usepackage[fleqn]{amsmath}
\usepackage{subfig}
\usepackage{eufrak}
\usepackage{mathrsfs}
\usepackage{eucal}
\usepackage{amsmath}
\usepackage{fontenc}
\usepackage{txfonts}

\title{Clustering analysis of high-redshift Luminous Red Galaxies in
Stripe 82} \author[N. Nikoloudakis et al.] {N.
Nikoloudakis$^{1}$\thanks{E-mail: nikolaos.nikoloudakis@durham.ac.uk},
T. Shanks$^{1}$, U. Sawangwit$^{1,2}$\\ 
$^{1}$Physics Department, University of Durham, South Road, Durham, DH1 3LE, UK \\
$^{2}$National Astronomical Research Institute of Thailand, Chiang Mai, 50200, Thailand}

\begin{document} 

\pagerange{\pageref{firstpage}--\pageref{lastpage}} \pubyear{2012}

\maketitle 
\begin{abstract}
We present a clustering analysis of Luminous Red Galaxies (LRGs) in
Stripe 82 from the Sloan Digital Sky Survey (SDSS). We study the angular
two-point autocorrelation function, $w(\theta)$, of a selected sample of
over 130\,000 LRG candidates via colour-cut selections in $izK$ with the
K band coverage coming from UKIRT Infrared Deep Sky Survey (UKIDSS) LAS.
We have used the cross-correlation technique of Newman (2008) to
establish the redshift distribution of the LRGs. Cross-correlating them
with SDSS quasi-stellar objects (QSOs), MegaZ-LRGs and DEEP2 galaxies,
implies an average redshift of the LRGs to be $z\approx1$ with space
density, \textbf{$n_{g}\approx3.20\pm0.16\times10^{-4}$ \rm h$^3$Mpc$^{-3}$}. For $\theta
\le 10'$ (corresponding to $\approx 10$ h$^{-1}$Mpc), the LRG $w(\theta)$
significantly deviates from a conventional single power-law as noted by
previous clustering studies of highly biased and luminous galaxies. A
double power-law with a break at $r_b\approx2.4$h$^{-1}$Mpc fits the
data better, with best-fit scale length, $r_{0,1}=7.63\pm 0.27$
h$^{-1}$Mpc and slope $\gamma_{1}=2.01\pm0.02$ at small scales and
$r_{0,2}=9.92\pm 0.40$ h$^{-1}$Mpc and $\gamma_{2}=1.64\pm0.04$ at large
scales. Due to the flat slope at large scales, we find that a standard
$\Lambda$ cold dark matter ($\Lambda \rm$CDM) linear model is accepted
only at $2-3\sigma$, with the best-fit bias factor, $b=2.74\pm0.07$. We
also fitted the halo occupation distribution (HOD) models to compare our
measurements with the predictions of the dark matter clustering. The
effective halo mass of Stripe 82 LRGs is estimated as $M_{\rm
eff}=3.3\pm0.6\times10^{13}$ h$^{-1}$M$_{\odot}$. But at large scales,
the current HOD models did not help explain the power excess in the
clustering signal.

We then compare the $w(\theta)$ results to the results of
\cite{Sawangwit11} from three samples of photometrically selected LRGs
at lower redshifts to measure clustering evolution. We find that a
long-lived model may be a poorer fit than at lower redshifts, although
this assumes that the Stripe 82 LRGs are luminosity-matched to the
$AA\rm \Omega$ LRGs. We find stronger evidence for evolution in the
{\it form} of the $z\approx1$ LRG correlation function with the above flat 2-halo slope
maintaining to $s\ga50$h$^{-1}$Mpc. Applying the cross-correlation test of \cite{ARoss11}, we find little evidence that the result is due to systematics. Otherwise it may represent evidence for primordial non-Gaussianity
in the density perturbations at early times, with $f^{\rm local}_{\rm
NL}=90\pm30$.

\end{abstract}


\section{Introduction} 
The statistical study of the clustering properties of massive galaxies 
provides important information about their formation and evolution which 
represent major questions for cosmology and astrophysics. The correlation
function of galaxies remains a simple yet powerful tool for implementing
such statistical clustering studies.
\citep[e.g.][]{Peebles80}.

A lot of interest has been concentrated specifically on measuring the
clustering correlation function of luminous red galaxies (LRGs)
\citep{Eisenstein01} \citep[see e.g][]{Zehavi05,Blake08,Ross08,Wake08,
Sawangwit11}. LRGs are predominantly red massive early-type galaxies, 
intrinsically luminous ($\ge 3 \rm L^{*}$) \citep{Eisenstein03,Loh06,Wake06} 
and thought to lie in the most massive dark matter haloes. They are also strongly biased
objects \citep{Padmanabhan07} and this coupled with their bright
luminosity makes their clustering easy to detect out to high redshifts.
For linear bias, the form of the LRG correlation function will trace
that of the mass but even in this case the rate of correlation function
evolution will depend on the bias model \citep[e.g.][]{Fry96}, which in
turn depends on the galaxy formation process.

The passive evolution of the LRG LF and slow evolution of the LRG
clustering \citep{Wake08,Sawangwit11} seen in SDSS, 2SLAQ and $AA\rm
\Omega$ Surveys already presents a challenge for hierarchical models of
galaxy formation as predicted for a cold dark matter (CDM) universe.
Since the LRG clustering evolution with redshift has been controversial,
a major goal is to use the angular correlation function to test if the
slow clustering evolution trend continues out to $z \approx 1$.

The uniformity of the LRG Spectral Energy Distributions (SEDs) with their 4000$\rm
\AA \ CaII \ H \& K$ break, offer the ability to apply a
colour-colour selection algorithm for our candidates. This technique
has been successfully demonstrated primarily by \citeauthor{Eisenstein01} in SDSS in
the analysis of LRG clustering at low redshift and then in 2SLAQ \citep{Cannon06}
and $AA\rm \Omega$ \citep{Ross08} LRG surveys at higher redshifts. For our study, the
available deep optical-IR $\textit{ugrizJHK}$ imaging data from the SDSS
+ UKIDSS LAS/DXS surveys in Stripe 82 will be used. This combination of
NIR and deep optical imaging data, on a moderate sample size of area $\sim
200$ deg$^2$, results in a sample of $\approx 130\,000$ LRG candidates at redshift 
$z \approx 1$.

The main tool for our clustering analysis will be the two-point angular
correlation function, $w(\theta)$, which has been frequently used in the
past, usually in cases where detailed redshift information was not
known. Hence, selecting Stripe 82 LRGs based on
colour-magnitude criteria, correspond to a rough photometric redshift
(photo-z) estimation based on the 4000$ \rm \AA$ break shifting through
the passbands. We shall apply the cross-correlation technique which was
introduced by \cite{Newman08} to measure the redshift distribution,
$n(z)$, of our photometrically selected samples. One of the main
advantages of $w(\theta)$ is that it only needs the $n(z)$ of the sample
and then through Limber's formula \citep{Limber53} it can be related
to the spatial two-point correlation function, $\xi(r)$.

In recent clustering studies, it was noted that the behaviour of
$\xi(r)$, which has previously been successfully described by a single
power-law of the form $\xi(r)=(r/r_{0})^{-\gamma}$, significantly
deviates from such a power-law at $\sim 1$ h$^{-1}$Mpc. The break in the
power-law, can be interpreted in the framework of a halo model, as arising
from the transition between small scales (1-halo term) to larger than a
single halo scales (2-halo term). Currently, our theoretical
understanding of how galaxy clustering relates to the underlying dark
matter is provided by the halo occupation distribution model (HOD, see,
e.g \citealt{Jing98,Ma00,Peacock00,Seljak00,Scoccimarro01,Berlind02})
via dark matter halo bias and halo mass function. Furthermore, the
evolution of HOD can also give an insight into how certain galaxy
populations evolve over cosmic time
\citep{White07,Seo08,Wake08,Sawangwit11}.

The outline of this paper is as follows. In Section 2, we briefly
describe the SDSS and UKIDSS data used in this paper, while in Section 3
we describe the angular function correlation function estimators and
their statistical uncertainties. In Section 4, we estimate the redshift
distribution through cross-correlations and then present the correlation
results together with their power-law fits, $\Lambda \rm$CDM model and a
halo model in Section 5. Section 6 is devoted to interpretation of the
clustering evolution. In section 7, we explore potential systematic
errors that might affect the large scale clustering signal. We then
argue that, if real, an observed large-scale clustering excess may be
due to the scale-dependent bias caused by primordial non-Gaussianity and
compare our results to other previous works in Section 8. Finally, in
Section 9 we summarize and conclude our findings.

Throughout this paper, we use a flat $\Lambda$-dominated cosmology with
$\Omega_{m}=0.27$, $H_{0}=100h \ \rm kms^{-1}Mpc^{-1}$, h=0.7,
$\sigma_{8}=0.8$ and magnitudes are given in the AB system unless
otherwise stated.
\section{DATA}

\subsection{LRG sample selection} \label{sec:selection}

We perform a $K$-band selection of high redshift LRGs in Stripe 82 based
on the combined optical and IR imaging data, $\textit{ugrizJHK}$, from
SDSS DR7 \citep{Abazajian09} and UKIDSS LAS surveys
\citep{Lawrence07,Warren07}, respectively. In previous studies,
\textit{gri} and \textit{riz} colours have been used to select low to
medium redshift LRGs, such as SDSS \citep{Eisenstein01}, 2SLAQ
\citep{Cannon06} and AA$\Omega$ \citep{Ross08} LRGs surveys up to $z
\approx 0.7$. In this work we aim to study LRGs at $z \approx 1$, thus
we use the \textit{izK} colour magnitude limits for our selection in
order to sample the 4000$ \rm \AA \ CaII \ H \& K$ break of the LRGs'
SED as it moves across the photometric filters (Fukugita et al. 1996; Smith et
al. 2002) taking advantage of the NIR photometry coverage from
UKIDSS LAS. Coupling the UKIDSS LAS to $K_{Vega} \leq 18 $ with the SDSS
ugriz imaging to $i_{AB} < 22.5$ in Stripe 82 produces an unrivaled
combination of survey area and depth. Our selection criteria are :
\begin{equation}
\begin{array}{l}
\displaystyle \rm SDSS \ Best \ Imaging \\
\displaystyle z-K +0.9(i-z) \geq 1.8 ,  \ Pri \ A \sim 700deg^{-2} \\
\displaystyle z-K +0.9(i-z) \geq 2.3, \  Pri \ B \sim 240deg^{-2} \\
\displaystyle z-K-0.9(i-z)\geq - 0.2      \\
\displaystyle -0.5 \leq i-z \leq 1.7         \\
\displaystyle z-K\leq 4.0                       \\
\displaystyle 17.0 \leq K  \leq 18       \\
\displaystyle z \leq 22.0.                                    
\end{array}
 \label{eq:colour_1}
\end{equation}

 \begin{figure}
 \begin{center}
\includegraphics[scale=0.25]{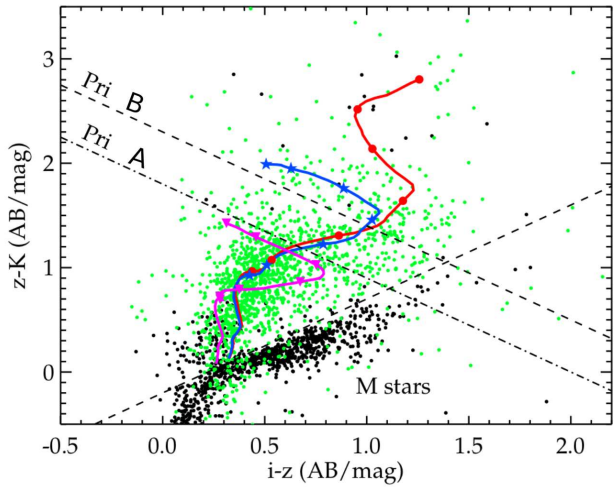}
 \end{center}
 \caption{\textit{iz} vs \textit{zK} colour-colour plot. Priority A and B correspond to the 
 $\sim700 \rm deg^{-2}$ and $\sim \rm 240deg^{-2}$ LRG samples, respectively. 
Objects with $J-K < 1.3$ which is typical for M stars are plotted as black circles where as 
those with $J-K \geq 1.3$ are plotted in green. 
Evolutionary tracks for single burst (red line) and $\tau =1 \rm Gyr$ (blue line) are overplotted from $z=0$ 
to 1.6 with symbols indicate $z$ interval of 0.2. The evolutionary track of late type galaxies 
(magenta line) is also shown for comparison.}
 \label{fig:izk}
 \end{figure}

The photometric selection of LRGs at $z>1$ requires a combination of
optical and NIR photometry as the $4000 \rm \AA $ band straddles the z
band. The selection of high-redshift LRGs is done on the basis of SDSS
$iz$ photometric data and the LAS $K$ band data (Fig.~\ref{fig:izk}).
LRG evolutionary models of \citet{Bruzual03} are overplotted for
single burst and $\tau =1 \rm Gyr$ galaxy models indicating the \textit{izk}
plane area where we should apply our selections in order to study the
high-z LRG candidates.

Late-type star contamination is a major problem in selecting a photometric sample of $z\approx1$ LRGs. Here the 
$z-K$ colour also helps to distinguish the M stars colour locus 
from those of galaxies. From Fig.\ref{fig:izk}, we see that most 
of the M stars lie at the bottom of the $izK$ colour plane. We identify 
these M stars by assuming their typical NIR colour, $J-K < 1.3$. 
However, this means that our selection criteria must involve $J$ band data 
and would reduce the sky coverage due to the data availability. Therefore 
we choose to exclude these M stars by applying a cut in $izK$ colour plane 
with the condition $z-K-0.9(i-z)\geq - 0.2$ in Eq.~\ref{eq:colour_1}.

All magnitudes and colours are given in SDSS \textit{AB} system and are
corrected for extinction using the Galactic dust map of
\citet{Schlegel98}. All colours described below refer to the differences
in `model' magnitudes \citep[see][for a review on model
magnitudes]{Lupton01} unless otherwise stated.

Applying the above selection criteria (Eq.~\ref{eq:colour_1}) on the SDSS DR7, we have two main LRG samples with a
total observed area (after masking) of $\approx200 \rm deg^2$. The first
sample has 130819 LRGs candidates with a sky surface density of $\approx
700 \rm deg^{-2}$ and the second one 44543 with a sky
density of $\approx 240 \rm deg^{-2}$. The $240\ \rm deg^{-2}$ LRG sample was
selected in such a way to check if the redshift distribution
implied by cross-correlations is higher than the $700\ \rm deg^{-2}$ LRG sample.

\section{THE 2-POINT ANGULAR CORRELATION FUNCTION MEASUREMENTS AND ERRORS}

\subsection{$w(\theta)$ Estimators}
 \label{sec:w_estim}

The probability of finding a galaxy within a solid angle $\delta\Omega$
on the celestial plane of the sky at a distance $\theta$ from a randomly
chosen object is given by\citep[e.g.][]{Peebles80}
 \begin{equation} 
\delta P =n[1+w(\theta)]\delta\Omega, 
\end{equation}
 where \textit{n} is the
mean number of objects per unit solid angle. The angular two-point
correlation function (2PCF) in our case, actually calculates the excess
probability of finding a galaxy compared to a uniform random point
process.

Different estimators can be used to calculate $w(\theta)$,
so to start with we use the minimum variance estimator from \citet{Landy93},

\begin{equation}
\textit{w}_{LS}(\theta)=1+\left ( \frac{N_{rd}(N_{rd}-1)}{N(N-1)}\right) \frac {DD(\theta)}{RR(\theta)}-2 \left (\frac{N_{rd}}{N}\right)\frac {DR(\theta)}{RR(\theta)}
\label{eq:ls_estim}
\end{equation}
 where $DD(\theta)$ is the number of
LRG-LRG pairs, $DR(\theta)$ and $RR(\theta)$ are the numbers of
LRG-random and random-random pairs, respectively with angular separation
$\theta$ summed over the entire survey area. $N_{rd}$ is the total
number of random points, $N$ is the total number of LRGs and $N_{rd}/N$
is the normalisation factor. For our calculation we used two LRG samples
(as explained in \S~\ref{sec:selection}) with different sky density,
thus the density of the random catalogue that we use is $\sim20$ times
and $\sim60$ times the number of the real galaxies for the first and
second LRG samples, respectively. Using a high number density random catalogue helps to ensure the
extra shot noise is reduced as much as possible.

We also compute $w(\theta)$ by using the \citet{Hamilton93} estimator
which does not depend on any normalisation and is given by,
\begin{equation}
\textit{w}_{HM}(\theta)=\frac{DD(\theta)\cdotp RR(\theta)}{DR(\theta)^2}-1.
\end{equation}
The Landy-Szalay estimator when used with our samples gives negligibly
different results to the Hamilton estimator. Note that the Landy-Szalay
estimator is used throughout this work except in \S \ref{subsec:gradient}
where we used both estimators to test for any possible gradient in
number density of our samples.

For the computation of the cross-correlations in \S \ref{sec:cross} and \S \ref{sec:tests} we use the estimator (\citet{Guo11}) :
\begin{equation}
\textit{w}_{cross}(\theta)=\frac{D_{G}D_{S}(\theta)-D_{G}R_{S}(\theta)-D_{S}R{D}(\theta)-R_{G}R_{S}(\theta)}{R_{G}R_{S}(\theta)}
\end{equation}
where the subscript $\textit{G}$ and $\textit{S}$ stands for the contribution in the pairs of the quantities that are cross-correlated in each case.

\subsection{Error Estimators}
 \label{sec:er_estim}

To determine statistical uncertainties in our methods, we used three different methods to estimate the errors on our measurements. 
Firstly, we calculated the error on $w(\theta)$ by using the Poisson estimate 
\begin{equation}
\sigma_{Poi}=\frac{1+w(\theta)}{\sqrt{ DD(\theta)}}.
\end{equation}

Secondly, we used the field-to-field error which is given by

\begin{equation}
 \sigma^{2}_{FtF}(\theta)= \frac{1}{N-1} \sum_{i=1}^{N}\frac{DR_{i}(\theta)}{DR(\theta)}[w_{i}(\theta)-w(\theta)]^{2},
\end{equation}
where N is the total number of subfields, $w_{i}(\theta)$ is an angular
correlation function estimated from the \textit{i}th subfield and
$w(\theta)$ is measured using the entire field. For this method we
divide our main sample to 36 subfields of equal size $\sim 6 \rm
deg^{2}$. We also reduce the number of subfields down to 18 with sizes
of $\sim 12 deg^{2}$ as we want to test how the results could deviate by
using different sets of subsamples. While Stripe 82 has only $\sim
2.5$deg height, our subfields with their $\sim 2.5$deg and $\sim5 $deg
widths are a reasonable size for estimating the correlation function up to
scales of $1-2$deg.

\begin{figure}
\begin{center}
 \includegraphics[scale=0.35,angle=90]{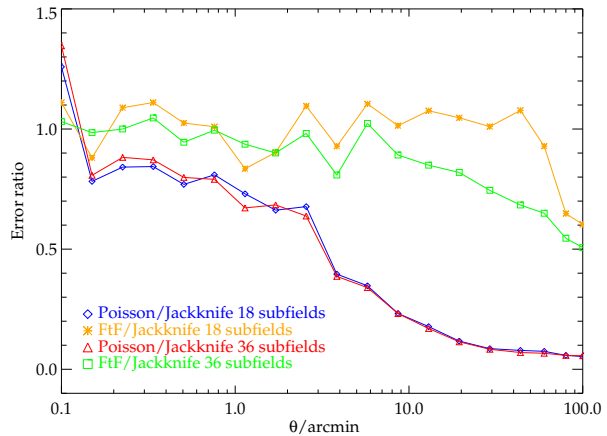}
 \end{center}
 \caption{Comparison of the measured error ratios of the Jackknife, field-to-field and the Poisson 
errors for the $w(\theta)$ measurements of the $700\ deg^{-2}$ Stripe 82 LRG sample. 
Two different resampling sets have been used for the Jackknife and field-to-field errors, the first one based on 36 subfields and the second from 18 subfields.}
 \label{fig:error_ratio}
 \end{figure}

Our final method is jackknife resampling, which is actually a bootstrap
method. This technique has been widely used in clustering analysis
studies with correlation functions (see, e.g \citealt{Scranton02,Zehavi05a,Ross07,Norberg09,Sawangwit11}).
The jackknife errors are computed using the deviation of the
$w(\theta)$ measured from the combined 35 subfields out of the 36
subfields (or 17 out of 18 when 18 subfields are used). The subfields
are the same as used for the estimation of the \textit {field-to-field}
error above. $w(\theta)$ is calculated repeatedly, each time leaving out
a different subfield and hence results in a total 36 (or 18)
measurements. The jackknife error is then

\begin{equation}
  \sigma^{2}_{JK}(\theta)= \sum_{i'=1}^{N}\frac{DR_{i'}(\theta)}{DR(\theta)}[w_{i'}(\theta)-w(\theta)]^{2},
\end{equation}
where $w_{i'}(\theta)$ is a measurement
using the whole sample except the \textit{i}th subfield and $
DR_{i'}(\theta) / DR(\theta)$ is approximately 35/36 (or 17/18) with
slight variation depending on the size of resampling field. A comparison of the error estimators can be seen in 
Fig.~\ref{fig:error_ratio}. Poisson errors are found to be much smaller compared to jackknife errors particularly at larger scales. Field-to-field errors give similar results
as jackknife errors, except at $\theta \gtrsim 10'$ where the FtF errors underestimate the true error due to missing cross-field pairs. Since 
the jackknife errors are better at a scale of order $100'$ which are of prime interest here, these are the error estimators that will be used in this work unless otherwise stated.

When calculated in small survey areas, $w(\theta)$ can be affected be an
\textquoteleft integral constraint\textquoteright , $ic$. Normally
$w(\theta)$ has a positive signal at small scales and if the surveyed
area is sufficiently small, this will cause a negative bias in
$w(\theta)$ at largest scales \citep{Groth77}, i.e. 
$w_{est}(\theta)=w(\theta)- ic$. The integral constraint can be
calculated from (see e.g. \citealt{Roche99}):

\begin{equation}
ic=\frac{\sum RR(\theta) w_{model}(\theta)}{\sum RR(\theta)},
\end{equation}
where for the $w_{model}(\theta)$ we assume
the standard $\Lambda\textrm{CDM}$ model in the linear regime
(\S\ref{sec:lcdm_model}). No integral constraint is initially applied to our
full sample results as the expected magnitude of $ic$ is smaller than the $w(\theta)$ amplitudes at scales analysed
in this paper. This position will be reviewed when we move on to discuss models with excess power at large scales in \S\ref{sec:tests}.

\begin{figure}
\begin{center}
 \includegraphics[scale=0.45]{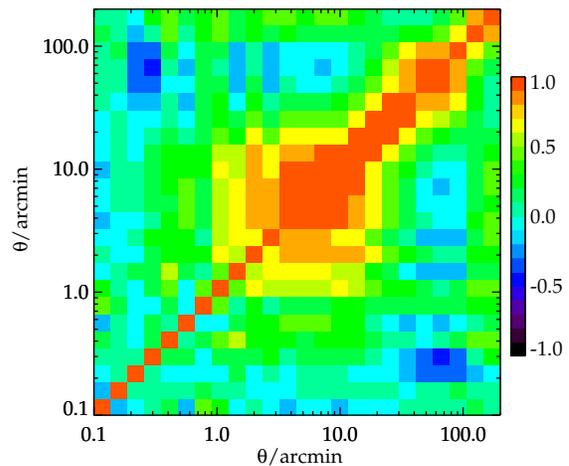}
 \end{center}
 \caption{The correlation coefficients $\mathbfss{r}_{ij}$, showing the level of correlation between each 
 angular separation bin for the $700\ \rm deg^{-2}$ Stripe 82 LRG sample as calculated by using 36 subfields.}
 \label{fig:36covar}
 \end{figure}

To provide robust and accurate results from the correlation functions, we are also interested in model fitting to the observed
$w(\theta)$ (see in \S\ref{sec:plaw}, \S\ref{sec:hod} and \S\ref{sec:lcdm_model}). Hence, for model fitting we will use the covariance matrix, which is calculated by:
\begin{equation}
\begin{array}{l}
\mathbf{C}_{ij}= \frac{N-1}{N} \sum_{i,j=1}^{N} [w(\theta_{i})- \overline{w(\theta_{i})}] [w(\theta_{j})- \overline{w(\theta_{j})}] 
\end{array}
\end{equation}
where the $w_{i}(\theta_{i})$ is the correlation function measurement value excluding the $i^{th}$ subsample and the factor $N-1$ corrects from the fact that 
the realizations are not independent (\citealt{Myers07,Norberg09,ARoss10,Crocce11,Sawangwit11}). 
The jackknife errors are the square-root of the diagonal elements of the covariance matrix, so we can now calculate the correlation coefficient, which is defined in 
terms of the covariance,

\begin{equation}
\mathbfss{r}_{ij}=\frac{\mathbfss{C}_{ij}}{\sqrt{\mathbfss{C}_{ii}\cdot
\mathbfss {C}_{jj}}}
\label{eq:corre_coeff}
\end{equation}  
where $\sigma^{2}_{i}=C_{ii}$ (see Fig.~\ref{fig:36covar}). We can see that the bins are strongly correlated at large scales.
The covariance matrix is more stable when we use 36 Jackknife
subfields instead of 18, so we will use only the covariance matrix for the case of 36 subfields.

\subsection{Angular Mask and Random Catalogue}
 \label{sec:masks}
\begin{figure}
\begin{center}
\includegraphics[scale=0.27]{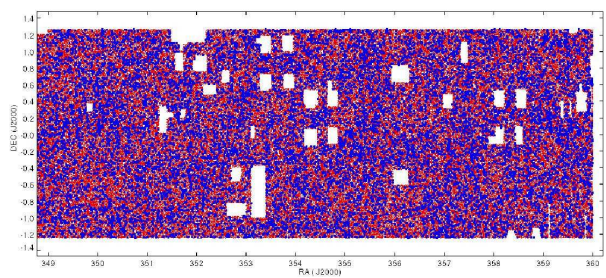}
\caption{A fraction of the total $\sim200\ \rm deg^{2}$ observed area in Stripe 82. LRG candidates (red) and 
random objects (blue), follow the same angular selection. Empty sky patches resulting from the lack of K-band 
coverage in the combined optical-IR data.}
\label{fig:stripe_82_area}
 \end{center}
\end{figure}

To measure the observed angular correlation function we must compare the actual
galaxy distribution with a catalogue of randomly distributed points.
The random catalogue must follow the same geometry as the real galaxy
catalogue, so for this reason we apply the same angular mask. The mask
is constructed from \textquoteleft BEST\textquoteright\ DR7 imaging sky
coverage\footnote{{\tt http://www.sdss.org/dr7}}. Furthermore, regions
excluded in the quality holes defined as \textquoteleft
BLEEDING\textquoteright,\,\ \textquoteleft TRAIL \textquoteright,\
\textquoteleft BRIGHT\_STAR \textquoteright \ and \textquoteleft
HOLE\textquoteright. The majority of the holes in the angular mask is
from the lack of $\textit K$ coverage in Stripe 82. The final mask is
applied to both our data and random catalogue (see
Fig.~\ref{fig:stripe_82_area}).

For generating the randomly distributed galaxies/points, we tried two
different ways in order to modulate the surface density of the random
points to follow the number density and the selection function of the
real data. The selection function of the random catalogue mimics only
the angular selection of the real data.

For the first method, we use a uniform density for the random points across the Stripe 82 area, so the normalization
factor, $N_{rd}/N$, to be $\sim20$ and $\sim60$ for the
$700\ \rm deg^{-2}$ and the $240\ \rm deg^{-2}$ LRG samples, respectively. A second
random catalogue was created by dividing Stripe 82 into six smaller
subfields ($15\times 2.5 \rm deg^2$ each) and normalizing the density of random points to the 
density of galaxies within each subfield. The
difference between the measured angular correlation function when we use
the \textquoteleft global\textquoteright\ or the \textquoteleft
local\textquoteright\ random catalogue is negligible.
We will use the \textquoteleft global\textquoteright\ random catalogue for the clustering analysis. 
A $k$d-trees code \citep{Moore01} has been used to
minimise the computation time required in the pair counting procedure.

\begin{figure}
\begin{center}
\includegraphics[scale=0.36,angle=90]{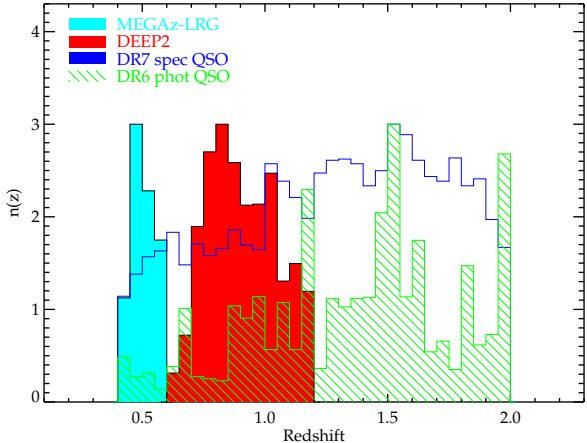}
\end{center}
\caption{Normalised redshift distributions of MEGAz-LRGs, DEEP2 galaxies and SDSS QSOs in Stripe 82 that are used in the cross-correlations with the LRG samples.}
\label{fig:nz}
\end{figure}

\section{LRG N(z) via Cross-Correlations}
 \label{sec:cross}

Even if the redshift of individual galaxies is not available, the
3-D clustering information can yet be recovered if the sample's redshift
distribution, \textit{n(z)}, is known. This can be achieved using
Limber's inversion equation \citep{Limber53} which can project the
spatial galaxy correlation function, $\xi(r)$, to the angular
correlation function given the $\textit{n(z)}$ of the sample:
\begin{equation}
w(\theta)=2\int_{0}^{\infty}dx f(x)^{2}\int_{0}^{\infty}du \ \xi(r=(u^{2}+x^{2}\theta^{2})^{1/2})
\label{eq:limbers}
\end{equation}
where $\textit{f(x)}$ is the galaxy
redshift selection function. For our photometric selected LRG samples,
only a very small fraction has a measured redshift, thus it is vital to
estimate the $\textit{n(z)}$ of the Stripe 82 LRG samples.

One method for estimating the redshift distribution of the sample could
be based on the various popular programs that derive photometric
redshifts (photo-z's). Photo-z estimates are based on the deep
multi-band photometry coverage, and work by tracing some specific
spectral features across the combination of filters which are then
compared with different type of objects SED templates. Indeed, our $\textit{izK}$ selection is a rough photo-z cut as we follow
the movement of the $4000 \rm \AA$ break across the selected bands. In
order to use the angular correlation function and the information that
is encoded we need the $\textit{n(z)}$ of our sample, hence we follow
the technique of \citet{Newman08} for reconstructing the LRG redshift
distribution from cross-correlations.

\subsection{Redshift distribution reconstruction} \
\label{sec:nz}

We employ Newman's method, which is about determining the underlying
redshift distribution of a sample of objects (LRGs in our case) through
cross-correlation with a sample of known redshift distribution. By
cross-correlating the sample (or samples) with known redshift and the
sample under consideration, if both samples lie at the same distance,
this will give a strong clustering signal. If the two samples that we
are cross-correlating are separated and are at different $z$ 
distances, no cross-correlation signal will result. Thus, through the
cross-correlations we can infer our photometrically selected LRG sample
$\textit{z}$ ranges.

Following \citet{Newman08} the probability distribution function of the redshift of the Stripe 82 LRG samples,
$\phi_{p}(z)$, is: 
\begin{equation} 
\phi_{p}(z)=w(z)\frac{3-\gamma}{2\pi}\frac{d_{A}(z)^{2}dl/dz}{ H(\gamma)r_{0,sp}^{\gamma}r_{max}^{3-\gamma}}
\label{eq:newmaneq}
\end{equation}
where $w(z)$ is the integrated cross correlation function,
$w_{sp}(\theta,z)$, of the LRG photometric samples with the samples of known
spectroscopic redshift (see \S\ref{ss:cross}), $H(\gamma)=\Gamma(1/2)
\Gamma((\gamma-1)/2)/ \Gamma(\gamma/2) $ where $\Gamma(\chi)$ is the
Gamma function, $d_{A}$ is the comoving angular distance and $dl$ is the comoving distance at redshift z. The comoving distance $r_{max}$ corresponds to the
maximum angle at given redshift, which must be large enough to avoid
nonlinear biasing effects.

To derive $\phi_{p}(z)$ via Eq.~\ref{eq:newmaneq} we must estimate
$w_{sp}(\theta,z) \sim \phi_{p}(z) \ r_{0,sp}^{\gamma_{sp}}$, since the
angular size distance, $d_{A}(z)$ and the comoving distance $l(z)$ are
given by the assumed cosmology. Thus we now require only
knowledge of the $\gamma_{sp}$ and $r_{0,sp}$ parameters as function of
redshift. Fortunately under the assumption of linear biasing, the
cross-correlation of the two samples under consideration is the result
of the geometric mean of the autocorrelation functions of the samples,
i.e. $\xi_{sp}=(\xi_{ss}\xi_{pp})^{1\over2}$, hence we can use the
information provided by autocorrelation measurements for each sample to
break the degeneracy between correlation strength and redshift
distribution.

Newman investigates the effect of systematics such as: different
cosmologies, bias evolution, errors from the autocorrelation
measurements and field-to-field zero points variations in the final
redshift probability distribution result. These issues could be more
important in the case of future photometric surveys aimed at
placing constraints on the equation of dark energy.

\subsection{Cross-Correlation data sets} 
\label{ss:cross}
Newman's angular cross-correlation technique requires the use of a data
sample with known spectroscopic, or sufficiently accurate photometric,
redshifts. For this reason we use a variety of samples with confirmed
spectroscopic and photometric redshifts for the cross-correlations with
Stripe 82 LRGs. The data samples that we use are: DEEP2 DR3 galaxies \citep{Davis03,Davis07} ,
MegaZ-LRGs \citep{Collister07}, SDSS DR6 QSOs \citep{Richards09} and
SDSS DR7 QSOs \citep {Schneider10}. In Fig.~\ref{fig:nz} we show the
normalised redshift distributions of all the samples and in Table~\ref{tab:data_sets} we
present the number of objects in each redshift bin.

\begin{table*}
\caption{Number of objects in each separate redshift-bin used for the cross-correlations with Stripe 82 LRGs}
\begin{tabular}{ccccc}
\hline\hline
& \multicolumn{4}{c} {sample}\\
 &DEEP2&MegaZ-LRGs&DR6 Photometric Sample&DR7 Spectroscopic sample\\
redshift	 & &  & &  \\
\hline\hline

0.4 - 0.6   &     	-	&  30503		&436            &  456  \\
0.6 - 0.8   &  3152    &    - 		&695            &  526  \\
0.8 - 1.0   &  5512    &    -		&1199          &  547  \\
1.0 - 1.2   &  3620	&     -		&1630          &  729  \\
1.2 - 1.4   &   -		&     -		&1312          &  820  \\
1.4 - 1.6   &     	-	&     -		&2646          &  854  \\
1.6 - 1.8   &     	-	&     -		&1193          &  803  \\
1.8 - 2.0   &     	-	&     -		&1990          &  668 \\
\hline
\end{tabular}
\label{tab:data_sets}
\end{table*}

   \begin{figure*}
   \centering
   \subfloat[]{\label{fig:trydsf1}\includegraphics[scale=0.36,angle=90]{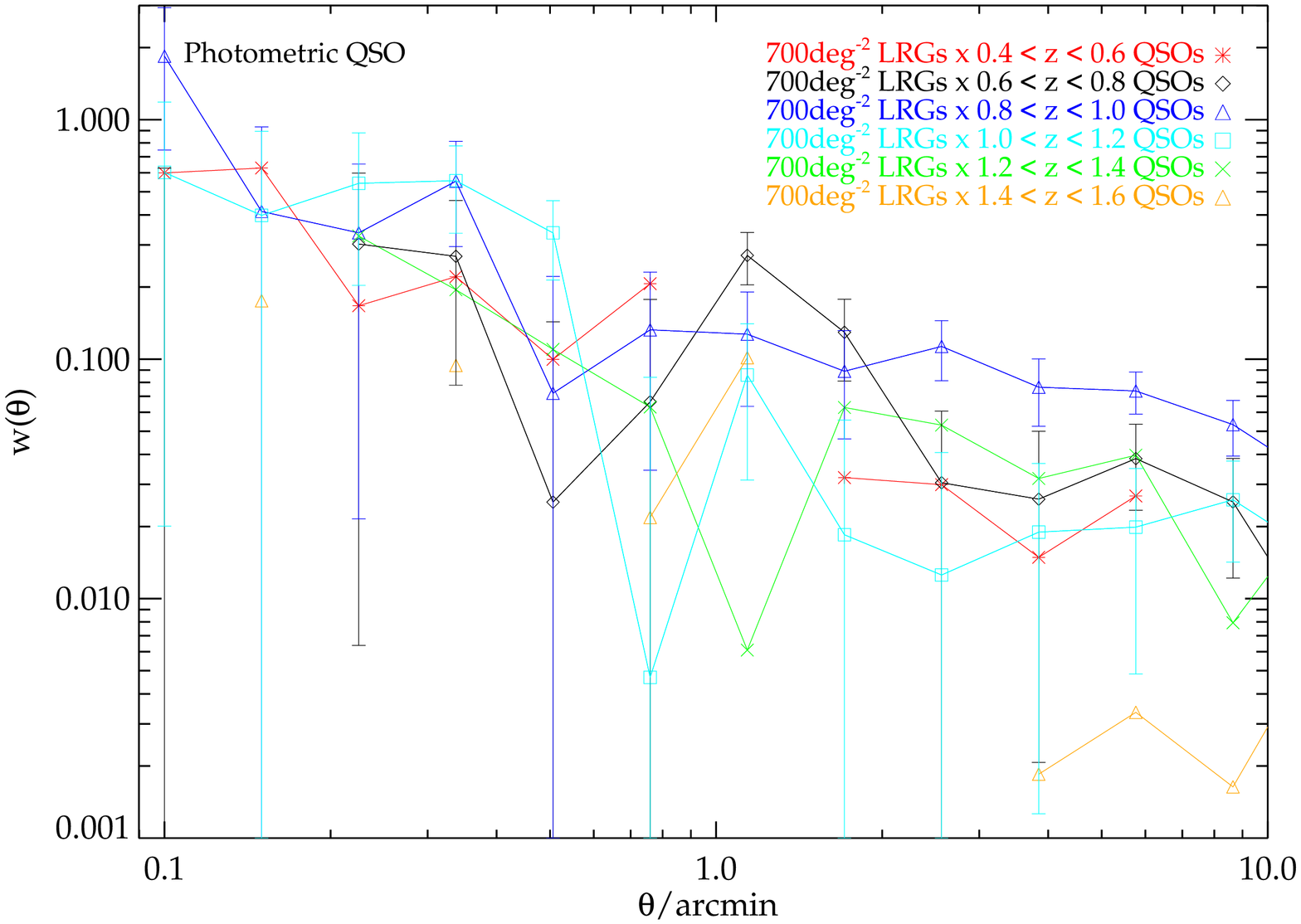}}                
   \subfloat[]{\label{fig:trsfsy2}\includegraphics[scale=0.36,angle=90]{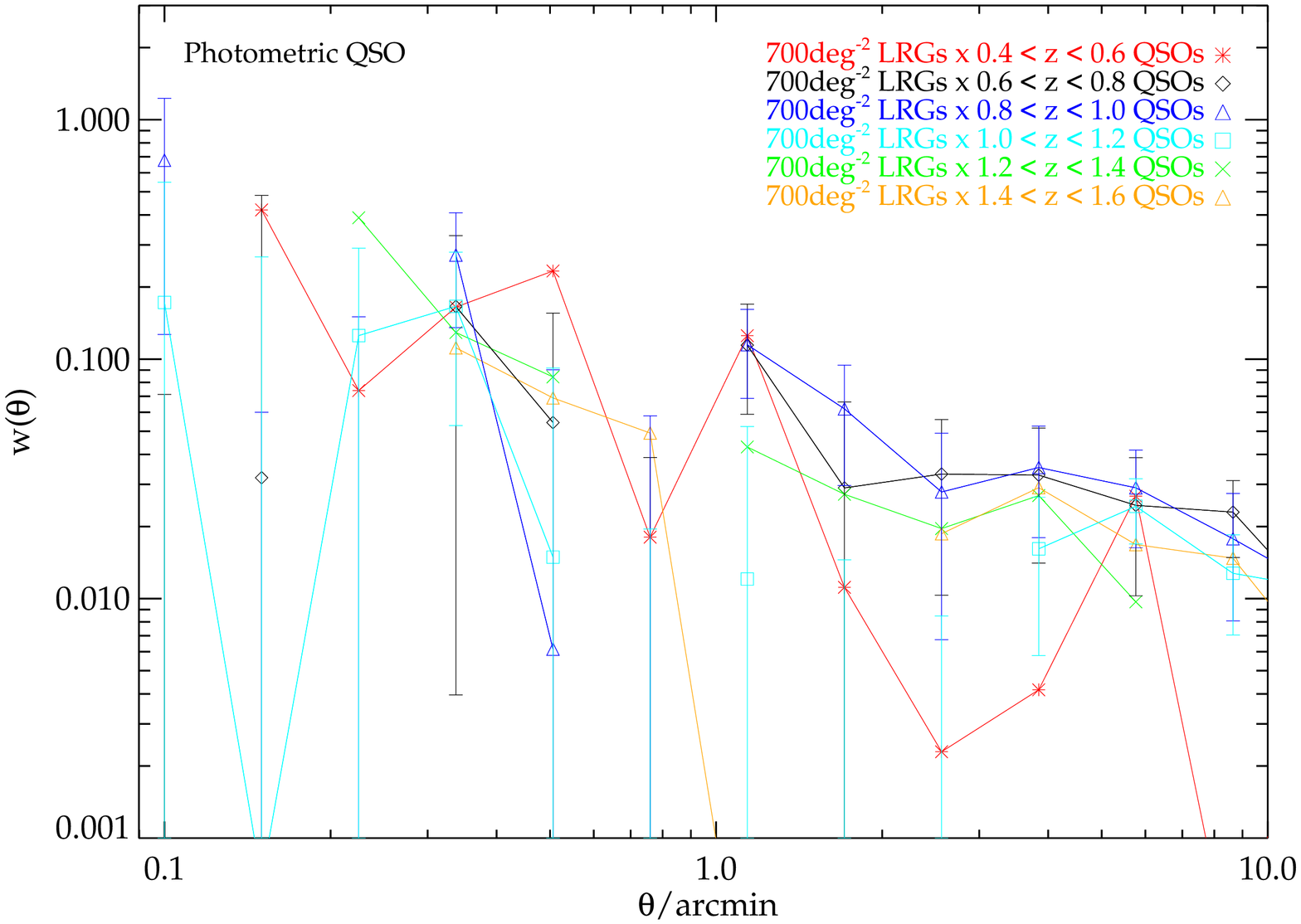}} 
   \caption{(a): Cross-correlation measurements of the $700\ \rm deg^{-2}$ Stripe 82 LRG sample with spectroscopic SDSS QSOs. (b): Same as (a) but now photometric SDSS QSOs are involved in the cross-correlations. Measurement uncertainties are $1\sigma$ jackknife errors.}
   \label{fig:QSO_cross}
  \end{figure*}

   \begin{figure*}
   \centering
   \subfloat[]{\label{fig:trydrtresf1}\includegraphics[scale=0.36,angle=90]{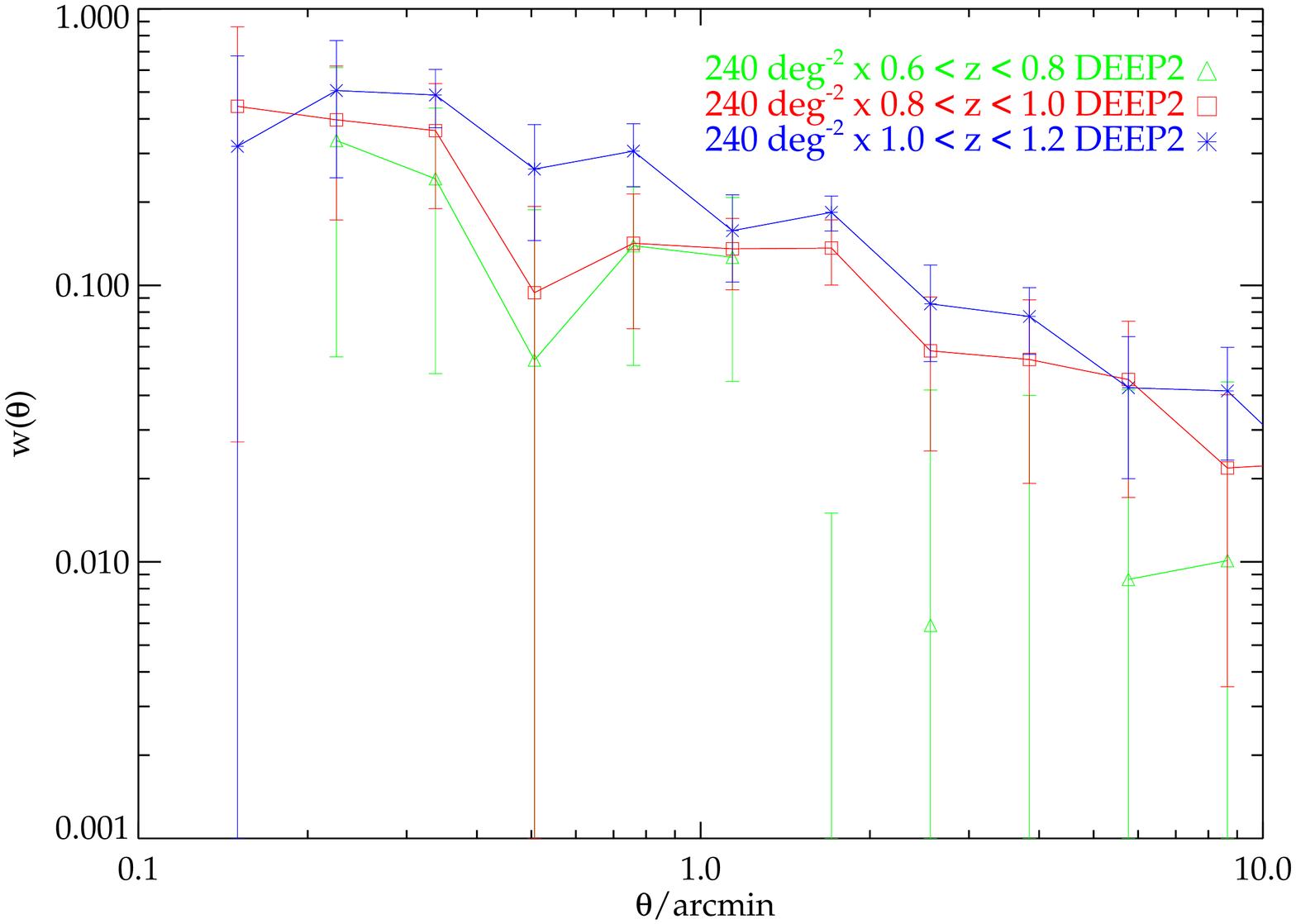}}                
   \subfloat[]{\label{fig:trsfsertery2}\includegraphics[scale=0.36,angle=90]{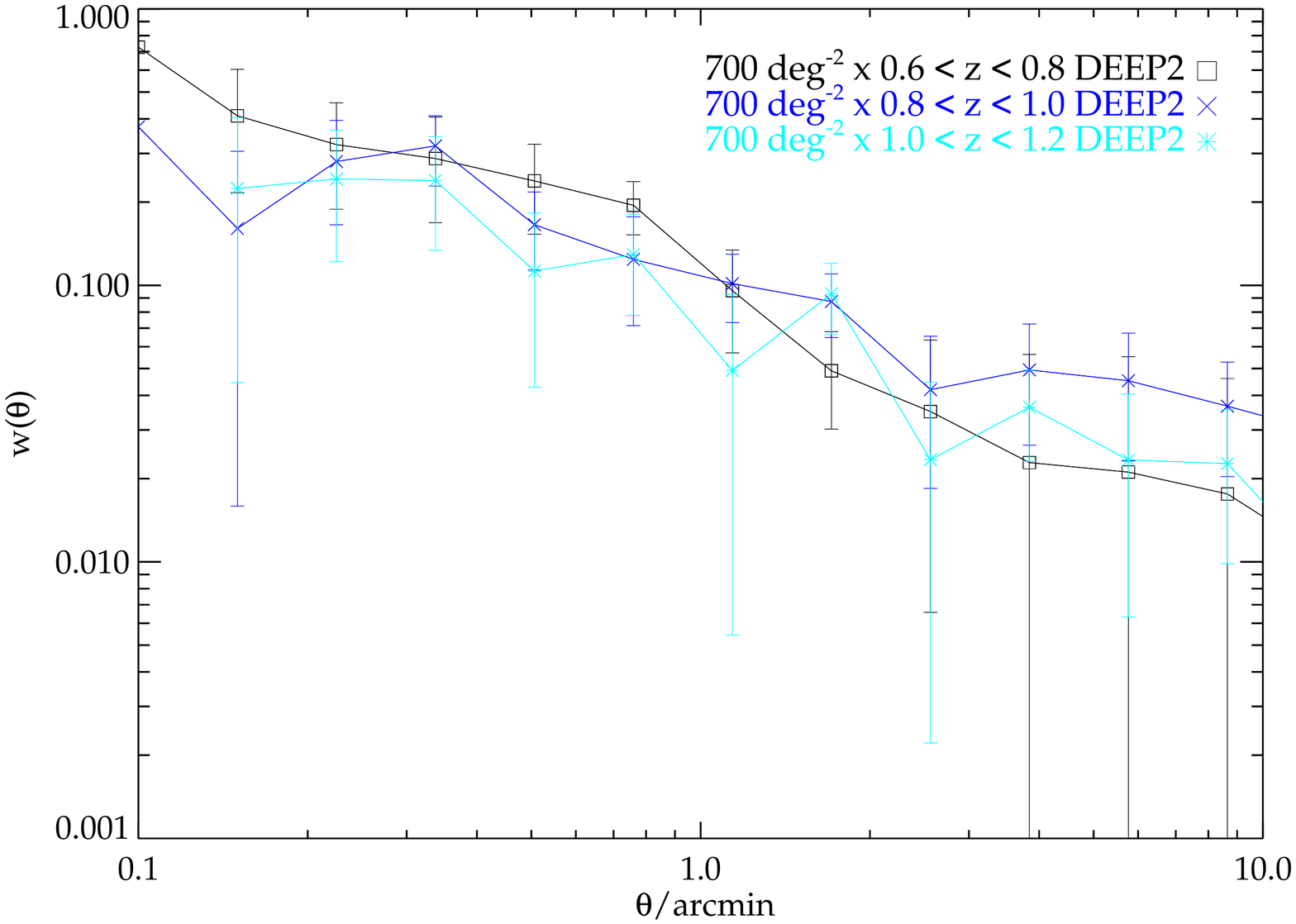}} 
   \caption{Cross-correlation measurements of the $240\ \rm deg^{-2}$ and $700\ \rm deg^{-2}$ Stripe 82 LRG samples with DEEP2 galaxies in (a) and (b), respectively. Uncertainties are $1\sigma$ jackknife errors.}
   \label{fig:DEEP2_cross}
  \end{figure*}

\begin{figure}
\begin{center}
\includegraphics[scale=0.36,angle=90]{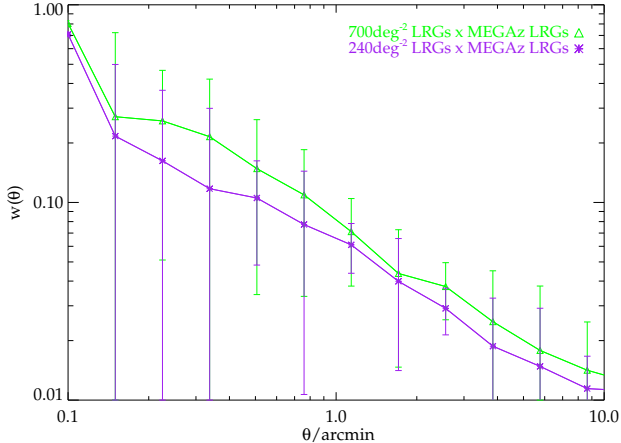}
\end{center}
\caption{Cross-correlation measurements of the $700\ \rm deg^{-2}$ (green diamond) and $240\ \rm deg^{-2}$ (purple star) Stripe 82 LRGs with MEGAz-LRGs, along with $1\sigma$ jackknife errors.}
\label{fig:MEGAz_corr}
\end{figure}

   \begin{figure*}
   \centering
   \subfloat[]{\label{fig:trydsf1a}\includegraphics[scale=0.33,angle=90]{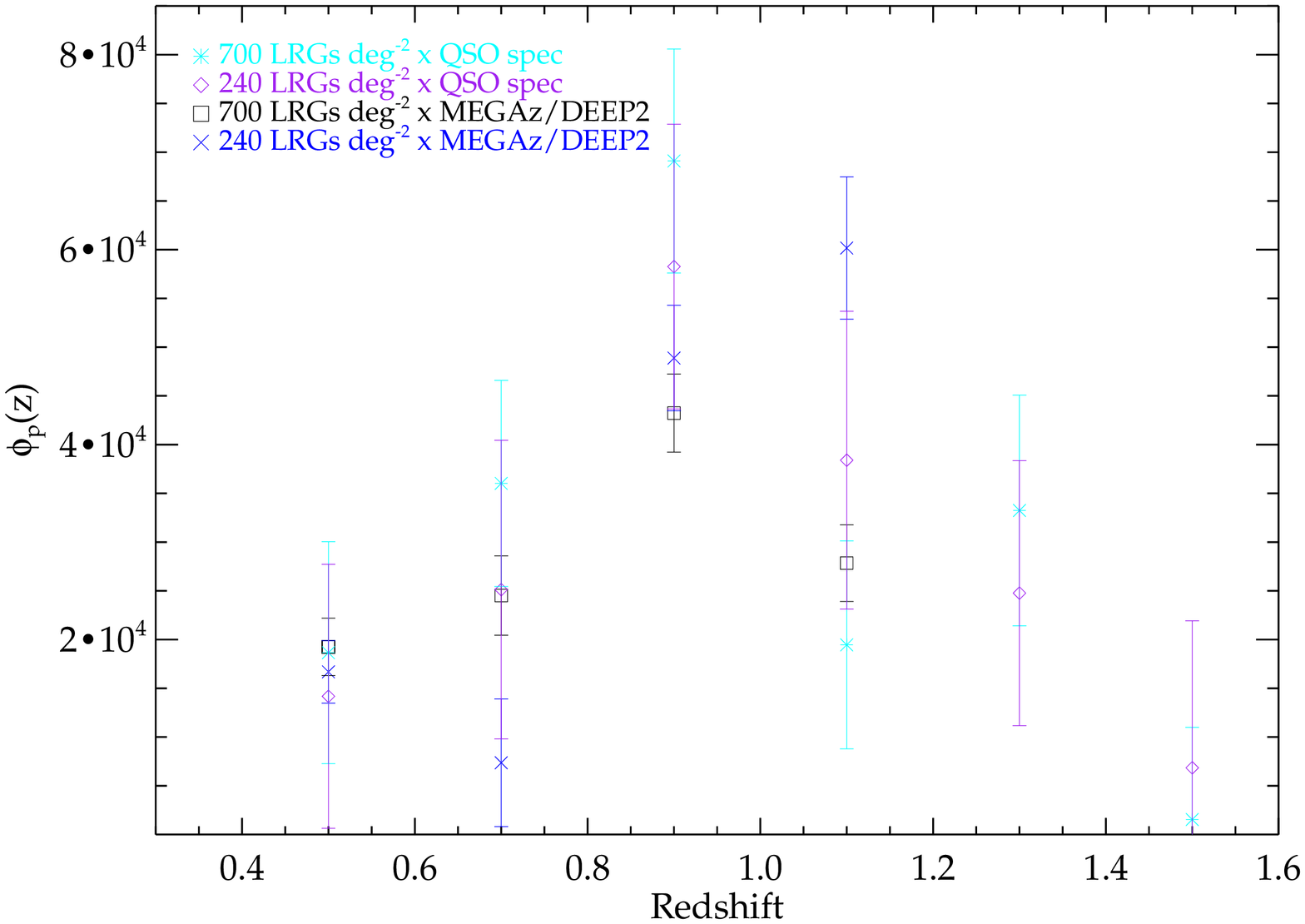}}                
   \subfloat[]{\label{fig:trsfsy2a}\includegraphics[scale=0.33,angle=90]{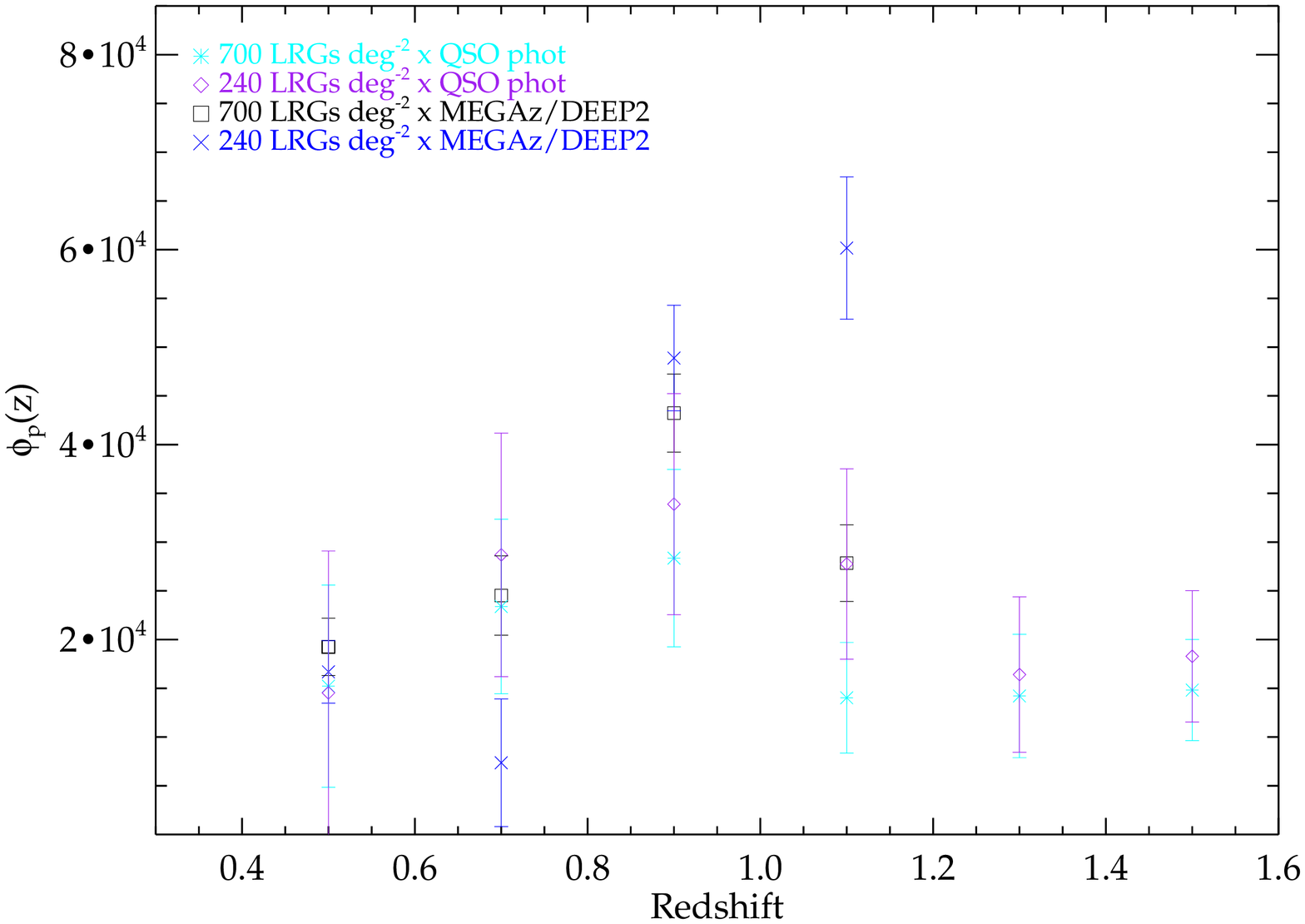}} 
   \caption{(a) The probability distribution function of the redshift, $\phi_{p} (z)$, of the $700\ \rm deg^{-2}$
 and $240\ \rm deg^{-2}$ Stripe 82 LRGs as estimated through cross-correlations with MEGAz-LRGs, DEEP2 galaxies and spectroscopic SDSS QSOs.
  (b) Same as in (a) but now using photometric SDSS QSOs instead of spectroscopic in the cross-correlations. Error bars shown in both cases are $1\sigma$ jackknife summed up to $6'$.}
   \label{fig:cross_nz1}
  \end{figure*}

By using the above data sets for cross-correlation we satisfy the
principal requirements of Newman's method, with the most important being
that the sky coverage of the data sets overlap the Stripe 82 LRGs.
It must be mentioned though that not all the redshift surveys have the
same sky coverage as Stripe 82 LRGs, so we reconstruct two redshift
distributions via the cross-correlations providing us with the
opportunity to check how much the $\textit{n(z)}$ cross-correlation
technique is affected by area selection. One $n(z)$ is reconstructed
by using all the data sets, the other $n(z)$ by using only SDSS QSOs in
the cross-correlations.

\subsubsection{SDSS DR6 $\&$ DR7 QSOs} QSO surveys are the main samples
that we used for our cross-correlation measurements and they span the
redshift range $0.4 \leq z \leq 2.0$. When we refer to QSO data sets, we
separate them into spectroscopic and photometric samples.

For the spectroscopic QSO sample we use the fifth edition of the SDSS
Quasar Catalog, which is based on the SDSS DR7 \citep{Schneider10}. The
original data set contains 105,783 spectroscopically confirmed QSOs,
from which only 5,403 in Stripe 82 have been used at $0.4 \leq z \leq
2.0$ for cross-correlations (Table~\ref{tab:data_sets}) with $i<
22$ ($\sim 28\%$ of QSOs at $i>20$).

The photometric QSO sample comes from the photometric imaging data of
the SDSS DR6 \citep{Richards09}. The parent catalogue contains $\sim
1,000,000$ QSOs candidates from which we use 11,101 with $i<21.3$ in
Stripe 82 and in the same redshift range as the spectroscopic QSOs.

In Fig.~\ref{fig:QSO_cross} we plot the cross-correlations between the
Stripe 82 LRGs and the SDSS QSOs. We show only the case for
cross-correlations of the $700\ \rm deg^{-2}$ Stripe 82 LRG sample with the
spectroscopic and photometric SDSS QSOs. Cross-correlation with the
$240\ \rm deg^{-2}$ LRG sample does not differ much. Errors shown here and for
the other cross-correlation cases are jackknife errors.

\subsubsection{DEEP2 Sample } The next sample of galaxies that we use is
DEEP2 DR3 galaxies \citep{Davis03,Davis07}. The survey coverage in
Stripe 82 is $\sim1.7 \ \rm deg^{2}$ with $i<24$. Galaxies in DEEP2 are split
in three redshift bins with 0.2 step in the redshift range $0.6 \leq z
\leq 1.2 $. The redshift distribution of the DEEP2 DR3 sample is shown
in Fig.~\ref{fig:nz}, with 12,284 galaxies in total. In Fig.~\ref{fig:DEEP2_cross} we show the
results of the cross-correlations of the $700\ \rm deg^{-2}$ and
$240\ \rm deg^{-2}$ LRG samples with the DEEP2 galaxies in the three
aforementioned redshift bins.\\

\subsubsection{MegaZ-LRG sample } The last sample that we use are LRGs
from the MegaZ-LRG photometric catalogue \citep{Collister07}. MegaZ-LRGs are
used only in the redshift range of $0.4 \leq z \leq 0.6 $ with $i<20$. This sample
offers us the ability to check the clustering properties of our
high-redshift LRG candidates with another sample of LRGs. The total
number of MEGAz LRGs that we use for cross-correlations is 30,503. In Fig.~\ref{fig:MEGAz_corr} are shown the cross-correlations between the
Stripe 82 LRGs and the MEGAz LRGs.

\subsection{Cross-Correlation results for $\textit{n(z)}$}

Having estimated the clustering signal from the cross-correlations of
the above samples, we proceed to the reconstruction of the redshift
distribution of the photometrically selected Stripe 82 LRG candidates.
To estimate the probability distribution function of the redshift,
$\phi_{p}(z)$, for the high-z LRG candidates we use equation
\eqref{eq:newmaneq}. The pair-weighted clustering signal of the
cross-correlations has been integrated up to $\approx6'$ for each
redshift bin.

In Fig.~\ref{fig:cross_nz1} we can see the two cases of the estimated
probability distribution function of the redshift for the high-z LRG
candidates. For the first case, $\phi_{p} (z)$ has been estimated by
using the spectroscopic SDSS QSOs whereas in the other case, $\phi_{p}(z)$ 
is estimated using only the photometric SDSS QSOs (DEEP2 galaxies and
MEGAz-LRGs are also always used). For both cases we plot the errors
estimated for each point in the redshift bin from the contributed
cross-correlated sample.

To estimate the redshift distribution, $\textit{n(z)}$, we use the
weighted mean for the $\phi_{p}(z)$ in each redshift bin, calculated
through :
\begin{equation}
n(z) = \frac{\displaystyle\sum_{i=1}^{k}(\phi_{p}(i)/\sigma_{i}^{2})}{\displaystyle\sum_{i=1}^{k}(1/\sigma_{i}^{2})},
\label{eq:w_weight}
\end{equation}
where $\textit{k}$ is the total number of bins at that redshift, $\phi_{p} (i)$
is the measured probability distribution function of each
cross-correlation data set in the \textit{i}th bin and $\sigma_{i}$ the
error on that measurement.

The spectroscopic QSO $\phi_{p} (z)$ in Fig.~\ref{fig:cross_nz1}a
compared to the photo-z case in Fig.~\ref{fig:cross_nz1}b, gives
increased probability at $z \sim 1$. This may be explained by the SDSS
QSO spectroscopic redshifts being more precise. For this reason, in our
analysis and in fitting models to our $w(\theta)$ results, we will use
only the spectroscopic $\textit{n(z)}$ for higher accuracy.

\begin{figure}
\begin{center}
\includegraphics[scale=0.33,angle=90]{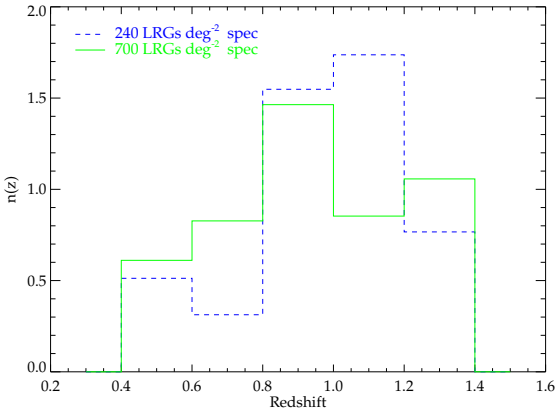}
\end{center}
\caption{Weighted normalised redshift distribution of the Stripe 82 LRGs
candidate samples when we use the spectroscopic SDSS QSOs along with the
DEEP2 and MEGAz-LRG data sets. As expected the $700\ \rm deg^{-2}$
sample (solid green line) n(z) peak is lower when compared with the
$240\ \rm deg^{-2}$ sample (dashed blue line).}
\label{fig:cross_nz2}
\end{figure}

In Fig.~\ref{fig:cross_nz2} we plot the normalized redshift
distribution of the $240\ \rm deg^{-2}$ and $700\ \rm deg^{-2}$ LRGs
samples as calculated from Eq.~\ref{eq:newmaneq} -~\ref{eq:w_weight}.
When we selected the two LRG samples from the $izK$ colour-plane, we
applied a redder selection for the $240\ \rm deg^{-2}$ sample (see
Eq.~\ref{eq:colour_1}), aiming for a sample with a slightly higher
redshift peak in the distribution as predicted from the evolutionary
tracks in Fig.~\ref{fig:izk}. This small difference may be seen
between the spectroscopic $\textit{n(z)}$ of the $700\ \rm deg^{-2}$ and
$240\ \rm deg^{-2}$ samples where the bluer cut has an average of $z
\sim 1$ where for the redder sample the average is $z \sim 1.1$. But
since the $700\ \rm deg^{-2}$ LRG sample has higher statistical accuracy
in the $\textit{n(z)}$ determination, the majority of our analysis will
be focused in this sample.

\section{RESULTS}

\subsection{Measured $w(\theta)$ and comparisons}
\label{sec:corre_compa}

\begin{figure}
\begin{center}
\includegraphics[scale=0.375,angle=90]{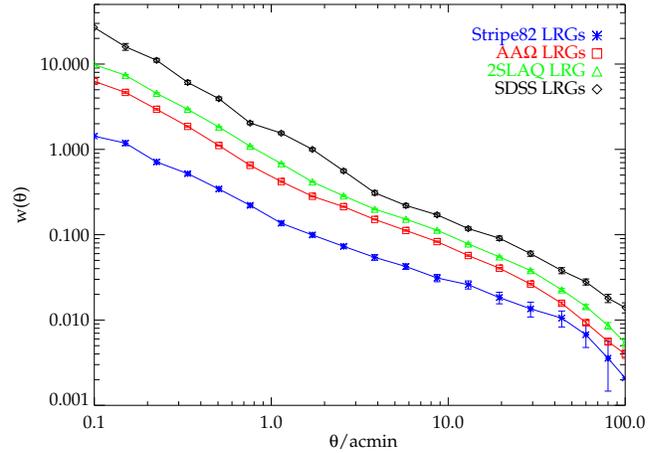}
\end{center}
\caption{The angular correlation function, $w(\theta)$, from the $700\ \rm deg^{-2}$ Stripe 82 LRGs (star), 
AA$\Omega$ LRGs (square), 2SLAQ LRGs (triangle) and SDSS LRGs (diamond). 
At small scales all of the measurements show similar clustering behaviour, 
but at large scales the Stripe 82 clustering slope appears to be flatter 
than the lower $z$ samples.}
\label{fig:wtheta}
\end{figure}

In Fig.~\ref{fig:wtheta} we compare the observed angular correlation
function of the $700\ \rm deg^{-2}$ LRG in Stripe 82 with \citet{Sawangwit11}
results. The $w(\theta)$ measurements are presented with 1$\sigma$
Jackknife errors.

The work of \citeauthor{Sawangwit11} involved three LRG data sets at $z \leq 1$ :
\begin{enumerate}
\item SDSS LRGs at $z\sim 0.35$
\item 2SLAQ LRGs $z\sim 0.55$
\item AA$\Omega$ LRGs $z\sim 0.68$
\end{enumerate}

From Fig.~\ref{fig:wtheta} we can see that at small scales,
$\theta\la1'$, the clustering trend for all the samples is similar but with
decreasing amplitude for increasing redshift. At larger scales, we
note that the $w(\theta)$ of the Stripe 82 LRGs seems to have a
flatter slope than the other samples, departing from the expected
behaviour for the correlation function.

Further comparisons below with the LRG clustering results of \citeauthor{Sawangwit11} 
will focus on the slope and amplitude of the $w(\theta)$ results,
with an initial view to interpret any changes in terms of evolution.
It is therefore of interest to see how the Stripe 82 sample match to the
LRG samples used in previous studies in terms of luminosity 
and comoving space density. 

A pair-weighted galaxy number density is given by \citep[see e.g.][]{ARoss09} :
\begin{equation}
 n_{g}=\int dz \frac{H(z)n(z)} {\Omega_{obs}\ c \ l^{2}(z)} \times n^{2}(z) \Big / \int dz n^{2}(z)
\label{eq:ndensity}
\end{equation}
where $\Omega_{obs}$ is the observed area of the sky, $l(z)$ is the comoving distance to redshift $z$ and $c$ is the speed of light. 
The observed space density for the 700deg$^{-2}$ 
Stripe 82 sample is found to be $\approx3.20 \pm 0.16 \times10^{-4} \rm h^3 Mpc^{-3}$. The quoted $1\sigma$ error has been estimated from the difference of the number density as calculated through Eq.~\ref{eq:ndensity} and by converting Fig.~\ref{fig:cross_nz2} into a plot of number density as a function of z (by dividing its bin by its corresponding volume).

Within the uncertainties of our $n(z)$, the 700deg$^{-2}$ sample appears
to have similar space density to that of the AA$\Omega$ LRG sample (see
Table~\ref{tab:plf} in \S \ref{sec:plaw}). However, in this study we do
not yet have redshift information for individual LRGs, not even for a subset
of the sample. Hence it is more uncertain if our sample has
similar luminosity as the LRG samples used by \cite{Sawangwit11}. We
therefore take the fact that the samples are number-density matched to
imply that they are also approximately luminosity matched which may turn
out to be a reasonable assumption (see e.g. \citealt{Sawangwit11}). This
then should enable us to compare the clustering slopes and amplitudes of
the AA$\Omega$ and Stripe 82 and infer any evolution independently of luminosity dependence.

\subsection{$w(\theta)$ and power-law fits} 
\label{sec:plaw}

Our first aim here is to fit power-laws to the Stripe 82 $w(\theta)$ to provide
a simple parameterisation of the results. Our second aim is to make
comparisons of the 3-D correlation amplitudes and slopes to measure
evolution. Both aims will require application of Limber's formula to
relate the 2-D and 3-D correlation functions.

We begin by noting that the simplest function fitted to correlation
functions is a single power-law with amplitude $r_0$ and slope $\gamma$.
In previous studies, the spatial correlation function has been
frequently described by a power-law of the form:

\begin{equation} \xi(r)=\left( \frac{r}{r_{0}}\right)^{-\gamma}.
\label{eq:xi}
 \end{equation} 
The angular correlation function as a
projection of $\xi(r)$ can be written as
$w(\theta)=\alpha\theta^{1-\gamma}$, commonly with a slope fixed at
$\gamma = 1.8$. The amplitude of the angular correlation function,
\textit{$\alpha$}, can be related with the correlation length $r_{0}$
through Limber's formula (Eq.~\ref{eq:limbers}) using the equation
\citep[][]{Blake08}:

\begin{equation} \alpha= C_{\gamma} \ r_{0}^{\gamma}\int dz \ n(z)^{2}
\left (\frac{dx}{dz} \right)^{-1} x(z)^{1-\gamma},
\label{eq:ampl}
\end{equation}
 where $n(z)$ is the redshift distribution, $x(z)$ is the comoving radial coordinate at
redshift z and the numerical factor 
$C_{\gamma} =\Gamma\left(\frac{1}{2}\right) \Gamma\left(\frac{\gamma}{2} -\frac{1}{2}\right) / \Gamma\left(\frac{\gamma}{2}\right)$.

\begin{table*}
\caption{Best fit parameters for the single and double power-law fits to the angular correlation function.}
\centering
\begin{tabular}{lccccccccccc}\\ \hline
\hline
Sample	 & $\bar{z}$& $n_{\rm{g}}$   & \multicolumn{3}{c} {Single power-law}   &	 \multicolumn{3}{c} {Double power-law }  \\
                 &        & $(h^3 Mpc^{-3})$&    $\gamma$& $r_{0}(h^{-1}\,\rm{Mpc})$&$\chi^2_{\rm{red}}$&
                  $\gamma_{1,2}$& $r_{0,1,2}(h^{-1}\,\rm{Mpc})$&$r_{b}(h^{-1}\,\rm{Mpc})$&$\chi^2_{\rm{red}}$\\
\hline
AA$\Omega$ LRGs	& 0.68 	&$2.7\times10^{-4}$	& $1.96\pm0.01$	& $7.56\pm0.03$	&42.8   	& $2.14\pm0.01$	& $ 5.96\pm0.03$&1.3	 & 3.4 \\
($110deg^{-2}$)	&          	&                                 	&                             	&                             	& 		& $1.81\pm0.02$	& $ 7.84\pm0.04 $&	 &  \\           
Stripe82 LRGs 		& 1.0  	& $3.20\pm 0.16\times10^{-4}$   	& $2.01 \pm 0.01$  	& $7.54\pm 0.16$	&5.89	& $2.01\pm 0.02$	&$7.63\pm0.27$&2.38	 &3.65   \\
($700\ deg^{-2}$)    &         	&                               	&                              &                             &	        & $1.64\pm0.04$     & $9.92\pm0.40 $&     &  \\ 
\hline \hline
\end{tabular}
\label{tab:plf}
\end{table*}

A deviation from a single power law at $\sim 1\rm h^{-1}Mpc$ has been
measured in previous studies \citep{Shanks83,Blake08,Ross08,Kim11,Sawangwit11}
and can be explained by the the 1-halo and 2-halo terms imprinted in the
clustering signal under the assumption of the halo model (see \S
\ref{sec:hod}). To parameterise the clustering characteristics of our
sample, we fit a single-power law and a double-power law 
to our measured angular correlation function. The double power-law form
is given as:

\begin{equation}
w_1(\theta) = \left(\frac{\theta}{\theta_{0,1}}\right)^{1-\gamma_1} (\theta<\theta_b)
\end{equation}
\begin{equation}
w_2(\theta) = \left(\frac{\theta}{\theta_{0,2}}\right)^{1-\gamma_2} (\theta \ge \theta_b)
\end{equation}
\noindent with $\theta_b$ to be the break point at $\approx 1.2'$ where the power-law slope changes from being steeper at
small scales ($<1.2'$), to flatter at large scales.

The power-laws are fitted in the range $0.1'<\theta<30'$ using the
$\chi^{2}$-minimization with the full covariance matrix constructed from
the jackknife resampling (see \S \ref{sec:er_estim}):
\begin{equation}
\chi^2= \sum_{i,j=1}^{N} \Delta w(\theta_{i})\mathbfss{C}_{ij}^{-1}\Delta
w(\theta_{j})
\label{eqn:chi2}
\end{equation}
where $N$ is the number of angular bins, $\Delta
w(\theta_{i})$ is the difference between the measured angular
correlation function and the model for the $i$th bin, and
$\mathbfss{C}_{ij}^{-1}$ is the inverse of the covariance matrix.

\begin{figure}
\begin{center}
\includegraphics[scale=0.36,angle=90]{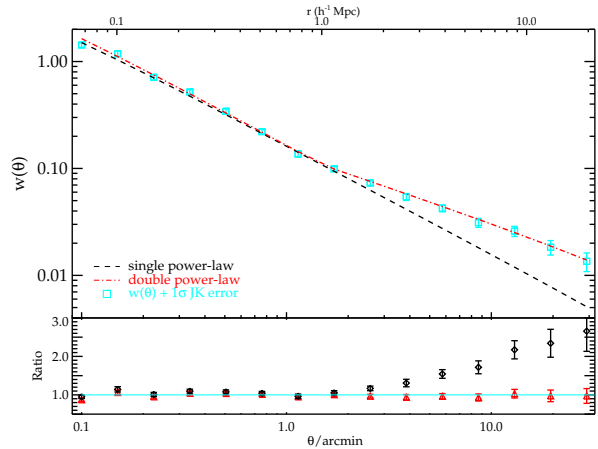}
\end{center}
\caption{The best-fit single power law (diamond) and double power law (triangle), for the $700\ \rm deg^{-2}$ LRGs candidates overplotted 
on the angular correlation function (square) with the $1\sigma$ Jackknife error. Lower panel shows the fitting residuals.}
\label{fig:plf}
\end{figure}

\begin{figure}
\begin{center}
\includegraphics[scale=0.45,angle=0]{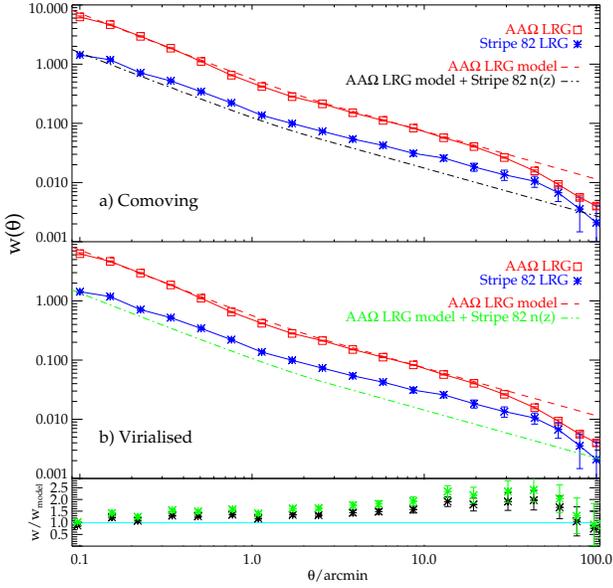}
\end{center}
\caption{\textbf{a)} 
The AA$\Omega$ LRG raw $w(\theta)$ measurements (red square) with 
predictions from comoving evolution model (dashed red line), using the best-fit double power-law $r_{0}-\gamma$ values 
with Limber's formula as \citet{Sawangwit11} calculated.
We then evolve the AA$\Omega$ best-fits utilising the estimated $700\ \rm deg^{-2}$ Stripe 82 LRG $\textit{n(z)}$ under the assumption of comoving evolution (dashed-dot black line) clustering. The observed Stripe 82 LRG $w(\theta$) is shown as well (blue star).\
\textbf{b)} Same raw measurements as above, but now compared to the virialised evolution clustering model. Stripe 82 LRG $w(\theta)$ measurements 
are described more accurate with comoving evolution at small and large scales compared to virialised evolution as it can been seen from the lower panel, where are plotted the residuals of the observed Stripe 82 $w(\theta)$ versus the comoving evolution (black star) and virialised evolution (green star) models, respectively.}
\label{fig:wmodel}
\end{figure}

For the single
power-law, our best-fit spatial clustering length and clustering slope
pair from Limber's formula are measured to be $r_{0}=7.54\pm 0.16 \rm h^{-1}Mpc$ and
$\gamma=2.01\pm0.01$ with associated reduced $\chi^{2}_{red}=5.89$.
The $r_{0}-\gamma$ pairs for the double power-law are $r_{0,1}=7.63\pm
0.27 \rm h^{-1}Mpc$ and $\gamma_{1}=2.01\pm0.02$ at small scales and
$r_{0,2}=9.92\pm 0.40 \rm h^{-1}Mpc$ and $\gamma_{2}=1.64\pm0.04$ at large
scales with a reduced $\chi^{2}_{red}=3.65$.
From the intersection of the 2 power law for $\xi(r)$, we have calculated the break scale, $r_{b}=2.38 \rm h^{-1}Mpc$. 
This is higher than the $r_{b}=1.3-2.2\rm h^{-1}Mpc$ estimated from the SDSS, 2SLAQ and $\textrm{AA}\Omega$ LRG surveys \citep{Sawangwit11}.

In Fig.~\ref{fig:plf} we
show the data points including the $1 \sigma $
Jackknife errors with the best-fitting power laws where the largest scale considered in the fitting was
$\theta<30'$, which corresponds to $r\la20h^{-1}Mpc$ at $z \sim1$ for the $700\ deg^{-2}$ LRG sample. 
Fig.~\ref{fig:plf} confirms that the double power-law clearly gives a
better fit to the data than the single power-law. Note that in the case
of the single power-law and the double power-law at small scales, our
results give \textit{$r_{0}-\gamma$} values consistent with outcomes
from previous studies. However, at large scales the Stripe 82 slope
($\gamma_2=1.64\pm0.04$) is significantly flatter than the
AA$\Omega$ result ($\gamma_2 = 1.81\pm0.02$).

Fig.~\ref{fig:wmodel} shows the double power-law fits for AA$\Omega$ (dashed red lines) taken 
from \citeauthor{Sawangwit11} and then evolved (black and green dot-dashed lines) to the Stripe 82 depth using 
Eq.~\ref{eq:ampl} under the assumptions of comoving and virialised clustering, respectively. We shall interpret the
amplitude scaling in the discussion of evolution in \S \ref{sec:clust_evol} later.
At this point we again note that the biggest discrepancy seems to be at
large scales where the Stripe 82 slope is increasingly too flat relative to
the AA$\Omega$ result. Fitted parameters are given in Table~\ref{tab:plf}, where 
the best-fit power-law parameters for the 
$\textrm{AA}\Omega$ LRG sample \citep{Sawangwit11} are also 
presented for comparison.

We note here that \citet{Kim11} studied the clustering of extreme red
objects (EROs) at $1<z<2$ in the SA22 field and they report a similar change of the
large scale slope. \citet{Gonzalez11} tried to fit clustering predictions
from semi-analytic simulations to the \citeauthor{Kim11} ERO $w(\theta)$ but
found that the model underpredicts the clustering at large scales.

\subsection{$\Lambda$CDM model fitting in the linear regime}
\label{sec:lcdm_model}

Since the standard $\Lambda$CDM model was found to give a good fit to the lower
redshift LRG samples of \citet{Sawangwit11}, we now check to see
whether the flatter large-scale slope of the Stripe 82 LRG $w(\theta)$ leads
to a statistically significant discrepancy with the $\Lambda \rm$CDM model at
$z\approx1$. 
We generate matter power spectra using the `CAMB' software \citep*{Lewis00}, including the case of non-linear growth of structure correction. For this reason 
we use the `HALOFIT' routine \citep{Smith03} in `CAMB'. Our models assume a $\Lambda$CDM Universe
with $\Omega_{\Lambda}=0.73$, $\Omega_{m}=0.27$, $f_{baryon}=0.167$, $\sigma_{8}=0.8$, $h=0.7$ and
$n_{s}=0.95$.
Then we transform the matter power spectra to
obtain the matter correlation function, $\xi_{\rm{m}}(r)$, using:
\begin{equation}
\xi_{\rm{m}}(r)=\frac{1}{2\pi^{2}}\int_{0}^{\infty}{P_m(k)k^{2}\frac{\
sin{kr}}{kr}}dk. 
\label{eq:pkxirtransform}
 \end{equation}

The relationship between the galaxy clustering and the underlying dark-matter clustering is given by the bias, $b_{g}$ : 
\begin{equation}
b_{\rm{g}}^2(r)=\frac{\xi_{\rm{g}}(r)}{\xi_{\rm{m}}(r)}. 
\end{equation}

\begin{figure}
\centering
\includegraphics[scale=0.365,angle=90]{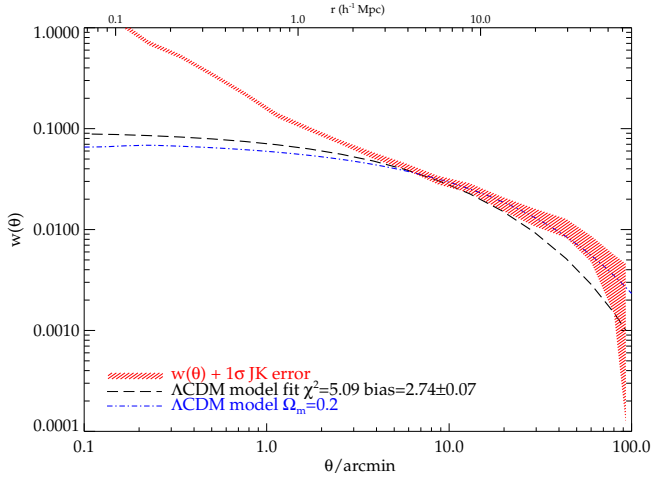}
\caption{The best-fit spatially flat, $\Lambda \rm$CDM model assuming
$\Omega_m=0.27$ compared to the observed $w(\theta)$ of Stripe 82 $700\
\rm deg^{-2}$ LRGs in the linear regime. The standard model cannot
explain the large scale power excess in the angular correlation function
of the Stripe 82 LRGs. The shaded area corresponds to $\pm1\sigma$
jackknife error. Also shown is a spatially flat $\Lambda \rm$CDM model
with the same parameters as before except for a lower value of
$\Omega_m=0.2$ and an arbitrary normalisation. The $\Omega_m=0.2$ model
appears to give a better fit than the standard $\Omega_m=0.27$ model.
}	
\label{fig:lcdm_model}
\end{figure}

As we are interested in the linear regime, we fit the
projected $\xi_{\rm{m}}(r)$ to the Stripe 82 LRG $w(\theta)$ in the range $4'\la\theta\la45'$,
corresponding to comoving separations $3\la r\la 30 \rm h^{-1}$Mpc. By
fitting the model predictions to the measured $w(\theta)$ it will result
with the best linear bias factor, the only free parameter in this case.
For our fitting, the $\chi^{2}$-minimization with the full covariance
matrix constructed from the jackknife resampling (see \S
\ref{sec:er_estim}) has been used.

The best-fit linear bias parameter is estimated to be $b=2.74\pm0.07$
with $\chi_{red}^{2}=5.09$. The upper limit of our fitted range in
$\theta$ was varied, while the lower limit stayed constant to avoid any
contribution from the non-linear regime. Thus, for the range $\sim
4'-30'$ the best-fit bias is $b=2.8\pm0.08$ with $\chi_{red}^{2}=4.72$
and at $\sim 4'-60'$ is $b=2.69\pm0.07$ with $\chi_{red}^{2}=5.18$. In
Fig. \ref{fig:lcdm_model} we plot the LRG $w(\theta)$ with the $1\sigma$
error and the $\Lambda \rm$CDM model with the best-fit bias. For low values
of the upper limit of the fitting range, the measured biases are in
approximate agreement with other results in the literature. But in terms
of the flat slope of $w(\theta)$ at large scales, the standard
$\Lambda \rm$CDM linear model is inconsistent with the data at the
$2-3\sigma$ level. One of the aims of the next section will be to see if
a HOD model can explain the flat large-scale slope of the $z\approx1$ Stripe 82 LRGs.

\subsection{Halo model analysis} \label{sec:hod}

We are going to use the approach of the halo model \citep[see][for a
review]{Cooray02} of galaxy clustering to finally fit our angular
correlation function results. Under the halo-model framework we can
examine the way the dark matter haloes 
are populated by galaxies through the Halo Occupation Distribution (HOD). Various
studies have used this model to fit their results
\citep[e.g.][]{Masjedi06,White07,Blake08,Wake08,Brown08,ARoss09,Zheng09,
Sawangwit11,Gonzalez11} as a way to explain the galaxy correlation
function and gain insight into their evolution. 
Specifically, we shall investigate whether the HOD model may be able to explain the flatter
slope of the correlation function observed here.

In the halo model, the clustering of galaxies is expressed by the contribution of number of pairs of galaxies
within the same dark matter halo (one-halo term, $\xi{1}$) and to pairs of galaxies in two separate haloes (two-halo term) :
\begin{equation}
\xi(r)=\xi_{1h}(r)+\xi_{2h}(r).
\label{eq:xi_hal}
\end{equation}
The 1-halo term dominates on small scales $\lesssim 1 Mpc$.

The fundamental ingridient in the HOD formalism of galaxy bias is the probability distribution
$P(N|M)$, for the number of galaxies $\textit{N}$ to hosted by a dark matter halo as a function of its mass $\textit{M}$.

We use the so-called centre-satellite three-parameter HOD model
\cite[e.g.][]{Seo08,Wake08,Sawangwit11} which distinguishes between the
central galaxy and the satellites in a halo. This separation has been shown in simulatations \citep{Kravtsov04} and has been commonly 
used in semi-analytic galaxy formation models in the last years \citep{Baugh06}.

Different HODs are applied for the central and satellite galaxies.
We assume that only haloes which host a central galaxy are able to host satellite galaxies.
The fraction of haloes of mass M with centrals is modelled as:
\begin{equation}
\left<N_{\rm c}|M\right>={\rm{exp}}\left(\frac{-M_{\rm{min}}}{M}\right).
\label{eq:Ncen} 
\end{equation} 
In such haloes, the number of satellite galaxies follows a Poisson distribution \citep{Kravtsov04} with mean:
\begin{equation}
\left<N_{\rm s}(M)\right>=\left(\frac{M}{M_1}\right)^\alpha.
\label{eq:Nsat}
\end{equation}
To describe the distribution of the satellite galaxies around the halo centre we use the NFW profile \citep{NFW97}.
So, the mean number of galaxies residing in a halo of mass $M$ is: 
\begin{equation}
\left<N|M\right>=\left<N_{\rm c}|M\right>\times(1+\left<N_{\rm s}|M\right>.
\label{eq:Ngal} 
\end{equation}
and the predicted galaxy number density from the HOD is then: 
\begin{equation}
n_{\rm{g}}=\int {\rm{d}}M\,n(M)\left<N|M\right>
\label{eq:ng}
\end{equation}
where $n(M)$ is the halo mass function, where in our case we use the model of \cite{Sheth99}.

From the HOD we can derive useful quantities which are
the central fraction :
 \begin{equation}
F_{\rm cen}= \frac{\int\rm{d}M\,n(M)\left<N_{\rm c}(M)\right>}{\rm{d}M\,n(M)\left<N_{\rm c}(M)\right>\left[1+\left<N_{s}(M)\right>\right]},
\label{eq:Fcent}
\end{equation}
and the satellite fraction of the galaxy population:
 \begin{equation}
F_{\rm sat}= \frac{1}{n_{\rm{g}}}\int {\rm{d}}M\,n(M)\left<N_{\rm c}(M)\right>\left<N_{\rm s}|M\right>,
\label{eq:Fsat}
\end{equation}
as $F_{sat}=1-F_{cen}$.
We can also determine the effective mass, $M_{eff}$, of the HOD: 
\begin{equation}
M_{\rm eff}= \frac{1}{n_{\rm{g}}}\int {\rm{d}}M\,n(M)M\left<N|M\right>,
\label{eq:Meff}
\end{equation}
and the effective large-scale bias:
\begin{equation}
b_{g}= \frac{1}{n_{\rm{g}}}\int {\rm{d}}M\,n(M)b(M)\left<N|M\right>,
\label{eq:blin}
\end{equation} 
where $b(M)$ is the halo bias, for which we use the ellipsoidal collapse model 
of \cite*{Sheth01}  and the improved parameters of \cite{Tinker05}.

\begin{table*}
\centering
\caption{Best-fit HOD parameters.}
\begin{tabular}{lccccccccc}

\hline
\hline
Sample	 & $\bar{z}$ & $M_{\rm min}$ & $M_1$& $\alpha$& $n_{\rm{g}}$   &  $M_{\rm eff}$ & $F_{\rm sat}$ & $b_{\rm lin}$ & $\chi^2_{\rm red}$  \\
       &    &($10^{13}h^{-1}M_{\odot}$)&($10^{13}h^{-1}M_{\odot}$)&  &$(10^{-4}h^3 Mpc^{-3})$&($10^{13}h^{-1}M_{\odot}$)& (per\,cent)& &   \\
          \hline
AA$\Omega$	& 0.68	& $1.02\pm0.03$ 	& $12.6\pm1.0$	& $1.50\pm0.03$	&$3.1\pm0.4$		&$3.0\pm0.1$& $9.0\pm0.09$ &$2.08\pm0.03$& 13.6 \\
Stripe82 $(10')$ 	& 1.0	& $3.09\pm0.75$ 	& $30.2\pm6.7$	& $2.38\pm0.12$	&$0.5\pm0.3$	&$4.0\pm0.6$& $2.13\pm1.0$ &$3.01\pm0.21$& 2.4 \\
Stripe82 $(30')$ 	& 1.0	& $2.57\pm0.31$ 	& $25.7\pm3.1$	& $2.28\pm0.04$	&$0.6\pm0.2$	&$3.6\pm0.5$& $2.62\pm0.07$ &$2.90\pm0.15$& 2.3 \\
Stripe82 $(45')$ 	& 1.0	& $2.19\pm0.63$ 	& $21.9\pm5.6$	& $2.24\pm0.12$	&$0.8\pm0.3$	&$3.3\pm0.6$& $3.17\pm0.10$ &$2.81\pm0.18$& 3.1 \\
Stripe82 $(60'$) 	& 1.0	& $2.19\pm0.21$ 	& $21.9\pm2.1$	& $2.25\pm0.05$	&$0.8\pm0.2$	&$3.3\pm0.3$& $3.17\pm0.08$ &$2.81\pm0.10$& 3.6 \\
        \hline
	\hline 
\end{tabular}
\label{tab:HODfit}
\end{table*}

   \begin{figure*}
   \centering
   \subfloat[]{\label{fig:gull}\includegraphics[scale=0.375,angle=90]{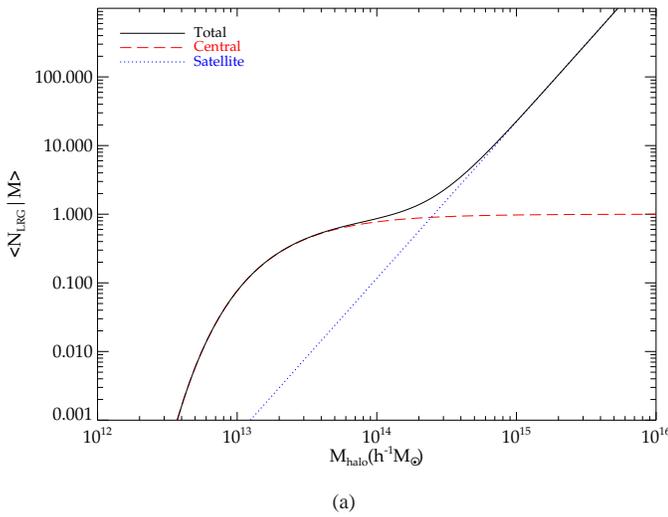}}                
   \subfloat[]{\label{fig:tiger}\includegraphics[scale=0.385,angle=90]{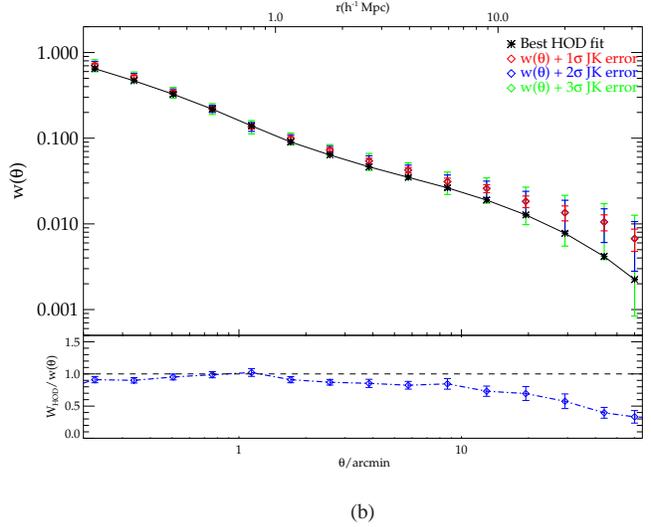}} 
\caption{(a) The mean number of LRGs per halo as a function of halo mass at $z=1$. The total, central and satellite contributions are shown by the solid, dashed and 
 dotted lines, respectively. (b) The measured angular correlation function $w(\theta)$ for the $700\ \rm deg^{-2}$ LRG sample with the best HOD fit (black star). 
The 1, 2 and $3\sigma$ Jackknife errors are shown in red, blue and green, respectively.}
   \label{fig:HOD}
  \end{figure*}

As the galaxy correlation function is the Fourier transform of the power spectrum, the 1-halo term and the 2-halo term of the clustering functions can be written as:
\begin{equation}
P(k)= P_{1{\rm h}}(k)+P_{2{\rm h}}(k).
\label{eq:PkHOD}
\end{equation}
Moreover the 1-halo term can be distinguished from the contribution of the central-satellite pairs, $P_{\rm cs}(k)$, and
satellite-satellite pairs, $P_{\rm ss}(k)$, \citep[see e.g.][]{Skibba09}:
\begin{equation}
P_{\rm cs}(k)= \frac{1}{n^2_{\rm g}}\int {\rm d}M\,n(M) 2\left<N_{\rm c}|M\right>\left<N_{\rm s}|M\right>u\left<k|M\right>,
\label{eq:Pkcs}
\end{equation}
and
\begin{equation}
P_{\rm ss}(k)= \frac{1}{n^2_{\rm g}}\int {\rm d}M\,n(M)\left<N_{\rm c}|M\right>\left<N_{\rm s}|M\right>^2u\left<k|M\right>^2,
\label{eq:Pkss}
\end{equation}
where $u\rm\left<k|M\right>$ is the NFW density profile in Fourier space and we have simplified 
the number of satellite-satellite pairs $\left<N_{\rm s}(N_{\rm s}-1)|M\right>$
to $\left<N_{\rm s}|M\right>^2$ since the satellites are Poisson-distributed.

The 2-halo term is evaluated as:
\begin{eqnarray}
P_{2{\rm h}}(k,r)&=&P_{\rm m}(k) \times \frac{1}{n_{\rm g}'^2} \nonumber \\
     & &\hspace{-14mm}\times\left[\int_0^{M_{\rm lim}(r)} \hspace{-2mm} {\rm d}M\,n(M)b(M,r)\left<N(M)\right>u(k,M)\right]^2,
\label{eq:Pk2h}
\end{eqnarray}
where $P_{\rm m}(k)$ is a non-linear matter power spectrum. We derive the mass limit,
$M_{\rm lim}(r)$, using the `$n_{\rm g}'$-matched' approximation of \citep{Tinker05},
which accounts the effect of halo exclusion: different haloes cannot overlap. 
$n_{\rm g}'$ is the restricted galaxy number density (Eq. B13 of \citet{Tinker05}). 

For the scale-dependent 
halo bias, $b(M,r)$, we use the model given by \cite{Tinker05}:
\begin{equation}
b^2(M,r)=b^2(M)\frac{\left[1+1.17\xi_{\rm m}(r)\right]^{1.49}}{\left[1+0.69\xi_{\rm m}(r)\right]^{2.09}},
\label{eq:bMr}
\end{equation}
where $\xi_{\rm m}(r)$ is the non-linear matter correlation function.
For the 2-halo term, we need to correct the galaxy pairs from the restricted galaxy 
density to the entire galaxy population.

By using Limber's formula
to project the predicted spatial galaxy correlation function $\xi(r)$ to
the angular correlation function $w(\theta)$ and we fit for a variety of
the three-parameter halo model ($M_{min}$, $M_{1}$, $\alpha$).

The best-fit model for each of our sample is then determined from 
the minimum value of the $\chi^{2}$-statistic using the full covariance matrix. 
We use the full covariance matrix over the range $0.25'<\theta<60'$ in our fitting.
Smaller scales are excluded in the fitting because any uncertainty in the $\xi(r)$ model 
can have a strong effect on $w(\theta)$ due to the projection.
To determine the $1\sigma$ error on the fits, the region of parameter space from the best fits with $\delta \chi^{2} \le 1$ ($1\sigma$ for 1 degree of freedom) is 
considered. For $b_{\rm lin}$, 
$M_{\rm eff}$, $F_{\rm sat}$ and $n_{\rm g}$ which depend on all the three main parameters, 
the considered region of the parameter space becomes $\delta \chi^2 \le 3.53$.

Fig. \ref{fig:HOD}a shows the resulting best-fit HOD of the mean number of LRGs per halo along with the central and satellite 
contributions. The best-fitting values for $M_{\rm min}$, $M_{1}$ and $\alpha$ where $M_{min}=2.19\pm0.63\times10^{13} \rm h^{-1}M_{\odot}$, $M_1=21.9\pm5.6\times10^{13} \rm h^{-1}M_{\odot}$ and $\alpha=2.24\pm0.12$, respectively. The 
associated values for $b_{\rm lin}$, $M_{\rm eff}$, $F_{\rm sat}$ and $n_{\rm g}$ are given in 
Table \ref{tab:HODfit}.

We see that the $\left<N|M\right>$ of the LRGs flatten at unity, as
expected from the assumption satellite galaxies are hosted by halos with
central galaxies. The LRGs as expected populate massive dark matter
haloes with the masses $\approx10^{13}-10^{14} \rm h^{-1} M_{\odot}$.
With the fraction of LRGs that are satellites being less than $5\%$, we
therefore find that $>95$\% of LRGs are central galaxies in their dark
matter haloes. The best fit linear bias, $b_{\rm lin} \approx 2.8$,
agrees with the prediction from \cite{Sawangwit11} in the case of a long
lived model for the LRGs and indicates that the LRGs are highly biased
tracers of the clustering pattern. The effective mass, $M_{\rm eff}
\approx 3\times10^{13} \rm h^{-1}M_{\odot}$, confirms that LRGs are
hosted by the most massive dark matter haloes. Despite the fact that we
use a higher redshift LRG sample, our best-fit HOD parameters are statistically not
too dissimilar to those found in previous LRG studies (eg see Table
\ref{tab:HODfit}).

In Fig. \ref{fig:HOD}b we show the best-fit model for $w(\theta)$,
compared to the data. The first thing we notice is that while at
small scales the best-fit HOD are in good agreement with the $w(\theta)$
measurements, at large scales the model fits only at $2-3\sigma$. The
flatter slope at large scales is responsible for that and we still are
not able to say if this can be explained by evolution in the linear
regime or any kind of systematic effect. In \S \ref{sec:tests} we will
check systematic errors that could affect our results.

Moreover, due to the high value of the best-fit reduced $\chi^{2}=3.1$,
we also try to fit the HOD models at different scales by using 4
different maximum $\theta $ bins of the covariance matrix in our fits,
which we present in Table \ref{tab:HODfit}. The fits at large scales did
not improve and above $45'$ there was not any change in the best-fit HOD
measurements.

Considering the two-halo term in the HOD model, one can see that the
bias in this regime is mostly scale-independent and the correction
factor is in fact having the opposite effect on the slope. The
scale-independent bias is simply the average of the halo bias, $b(M)$,
weighted by the halo mass function and the mean number of galaxies
hosted by the corresponding halo. One way to boost the large-scale
amplitude is to increase $M_{\rm min}$ and therefore increase the mass
range of the halo where most galaxies occupy and hence linear bias and
amplitude of the two-halo term. However, to compensate for the increase
numbers of satellite galaxies (and consequently small-scale clustering
amplitude) one must also increase $M_1$, the mass at which a halo hosts
one satellite galaxy on average. And in order to produce the overall
flatter slope one needs to increase $M_1/M_{\rm min}$. However, this
would still overpredict the clustering amplitude in the intermediate
scales, $r \sim5-10 \ \rm h^{-1}Mpc$. Note that our best-fit HOD gives
$M_1/M_{\rm min} \approx 10$, consistent with previous results for lower
redshift LRGs of \citep{Sawangwit11} and \citep{Wake08}. However, as
noted earlier including $w(\theta)$ bins at larger and larger scales
does not change the best-fit parameters which means that $M_1/M_{\rm
min}$ also remains unchanged due to the reason discussed above. We
therefore conclude that the HOD prescription in the framework of
standard $\Lambda \rm$CDM cannot explain the observed large-scale slope
in $w(\theta)$ of the $z\approx1$ LRG sample.

\section{Clustering Evolution}

\subsection{Intermediate scales}
\label{sec:clust_evol}
First, we compare the clustering of the $z\approx1$ Stripe 82 LRG sample
to the lower redshift $z\approx0.68$ $AA\Omega$ LRG sample. We recall
that these LRG samples have approximately the same space density and so
should be approximately comparable. We follow \citet{Sawangwit11} and by using our best-fit $r_{0}$ and $\gamma$
we make comparison with their data and models via the integrated
correlation function in a $20\ \rm h^{-1}Mpc$ sphere, $\xi_{20}$.

$AA\Omega$ LRG results are described better with the long-lived model of
\citet{Fry96}. Fry's model assumes no merging in the clustering
evolution of the galaxies while they move within the gravitational
potential, hence the comoving number density is kept constant. The bias
evolution in such a model is given by:
\begin{equation}
 b(z)=1+\frac{b(0)-1}{D(z)}
\end{equation}
where D(z) is the linear growth factor.

\begin{figure}
 \begin{center}
 \includegraphics[scale=0.35,angle=90]{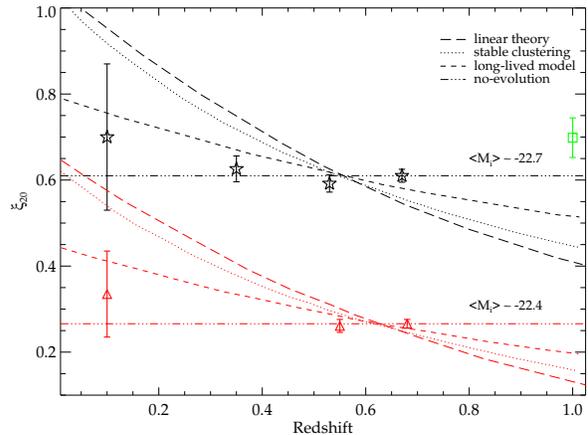}
 \end{center} 
 \caption{The LRG $\xi_{20}$ measurements as a function of redshift and luminosity from Sawangwit et al. (2011). Lowest redshift data are early-type galaxies from 
 2dFGRS (Norberg et al. 2002). Stars represent the brighter samples (SDSS, 2SLAQ* and $AA\Omega^*$-LRG), where the lower luminosity samples, triangles, have been lowered by 0.2 for clarity. The 700 deg$^{-2}$ Stripe 82 LRGs $\xi_{20}$ measurement is at $z=1$ (square). }
 \label{fig:xi20}
 \end{figure}

However, the flat slope beyond 1h$^{-1}$Mpc causes a highly significant,
$\approx50$\%, rise in $\xi_{20}$ above the $AA\Omega$ $\xi_{20}$ as we
can see in Fig.~\ref{fig:xi20} (see also Figs.~\ref{fig:wmodel}a,b). If
we assume that the 2 samples are matched then we would conclude that all
of the models discussed by \citet{Sawangwit11} were rejected.

One possibility is that the 700deg$^{-2}$ LRG sample is closer to the
SDSS and $AA\Omega^*$ LRG space density of $1.1\times10^{-4}$
h$^{-3}$Mpc$^{-3}$ because the LRG $\xi_{20}$ fits the extrapolated
models better there. If so, then this would imply that the Stripe 82 LRG
$n(z)$ width was underestimated in the cross-correlation procedure and
this would then increase the deprojected amplitude of $\xi(r)$,
suggesting that this explanation may not work. Similarly a larger
correction for stellar contamination would also produce a higher Stripe
82 clustering amplitude. We do not believe that looking further into the evolution of the bias \citep{Papageorgiou12}
and DMH is warranted until we understand the flat slope of the Stripe 82
$w(\theta)$ at large scales.

\subsection{Small scales}
At smaller scales ($r<1\ \rm h^{-1}Mpc$) the situation is less complicated by
the flat large-scale slope. Here \citeauthor{Sawangwit11} found that a
virialised model gave a better fit to the slightly faster evolution
needed to fit the small-scale correlation function amplitudes than a
comoving model. But in the present case, the scaling between the AA$\Omega$
and Stripe 82 LRGs in Fig.~\ref{fig:wmodel}a,b, shows that here the comoving model is
preferred at small scales over the faster virialised evolution. This
fits with the more general picture of the Stripe 82 LRGs presenting a
higher amplitude than expected all the way down to the smallest scales.
Unfortunately the remaining uncertainty in the Stripe 82 LRG luminosity
class is still too large to make definitive conclusions on this
evolution possible.

\subsubsection{HOD Evolution}
\label{sec:hodevol}

Given the uncertainty in $\xi_{20}$ caused by the flat $w(\theta)$ slope
on intermediate - large scales, we will extend further the studies at small-scales, using
the HOD model to interpret the small-scale clustering signal of the
LRGs. Based on the HOD fit at $z\approx1$, we again follow \cite{Sawangwit11}, (and references therein)
and test long-lived and merging models by comparing the predictions of 
these models to the SDSS HOD fit from \citeauthor{Sawangwit11}. These authors and also \cite{Wake08}
found that long-lived models were more strongly rejected at small scales ($r<1$ h$^{-1}$Mpc)
than at intermediate-large scales.

Again we follow the approach of \cite{Wake08}) and \cite{Sawangwit11} who assumed a form for 
the conditional halo mass function \cite{Sheth02} and a sub-Poisson distribution for the number 
of central galaxies in low-redshift haloes of mass $M$ such that
\begin{equation}
\left<N_{\rm c}(M)\right>=1-\left[1-\frac{C(M)}{N_{\rm max}}\right]^{N_{\rm max}}, 
\end{equation}
where $N_{\rm max}={\rm int}(M/M_{\rm min})$,
\begin{equation}
C(M)=\int_0^M {\rm d}m\,N(m,M)\left<N_{\rm c}(m)\right>
\end{equation}
and $N(m,M)$ is the expression of \cite{Sheth02} for the conditional halo mass which generalize those 
of \cite{Lacey93}.
The mean number of satellite galaxies in the 
low-redshift haloes is then given by
\begin{equation}
\left<N_{\rm c}(M)\right>\left<N_{\rm s}(M)\right>=S(M)+f_{\rm no-merge}\left[C(M)-\left<N_{\rm c}(M)\right>\right],
\label{eq:nomerge}
\end{equation}
where 
\begin{equation}
S(M)=\int_0^M {\rm d}m\,N(m,M)\left<N_{\rm c}(m)\right>\left<N_{\rm s}(m)\right>.
\end{equation}
and the main parameter is $f_{\rm no-merge}$ which is
the fraction of un-merged low-$z$ satellite galaxies which were high-$z$
central galaxies.

This model is called `central-central mergers' in
\cite{Wake08}. More massive high-z central galaxies are more likely to
merge with one another or the new central galaxy rather than
satellite-satellite mergers.

\begin{figure}
\hspace{-0.45cm}
\centering
  \includegraphics[scale=0.37,angle=90,]{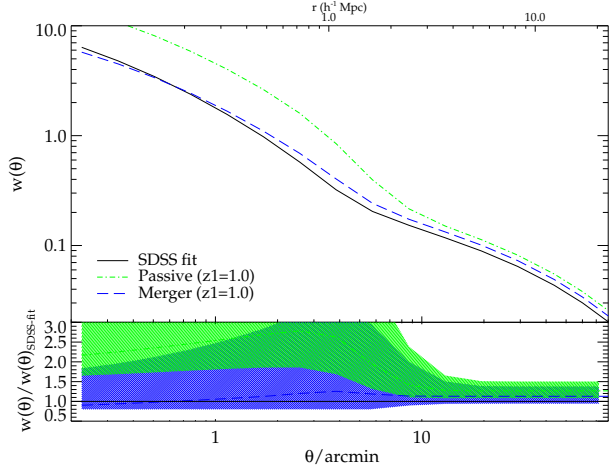}
	\caption{The predicted SDSS LRG $w(\theta)$'s at $z_{later}=0.35$ for the case of passively ($f_{\rm no-merge}=1$) evolving 
	the best-fit HOD of Stripe 82 LRGs sample from $z_{earlier}=1$ and the case where central galaxies merging is allowed from $z_{earlier}=1$ 
	($f_{\rm no-merge}=0.21$), in green dot-dashed line and blue long-dashed line, respectively.
 	The bottom panel shows the ratios of the evolved $w(\theta)$'s to the SDSS best-fit, the shaded regions 
	signify the $1\sigma$ uncertainties.}
	\label{fig:HODevolve}
\end{figure}

Setting $f_{\rm no-merge}=1$ means that there is no merging of initial
central galaxies in subsequently merged haloes, so it is similar to the
passive/long-lived model. $f_{\rm no-merge}$ equals to 0 means that all
the central galaxies in haloes at high redshift merge to form new
central and/or satellite galaxies in the low redshift haloes. In the
analysis below, we use the best-fit HOD model values as estimated for
scales up to $45'$ (see Table \ref{tab:HODfit}).

The $f_{\rm no-merge}=1$ case is shown as the $w(\theta)$ passive model
in Fig.~\ref{fig:HODevolve} and is clearly rejected by the data at $\theta\lesssim 10'$(see lower panel). Best-fit HOD predictions of the
satellite fraction in the case of the passively evolved LRGs from
$z_{earlier}=1$ to $z_{later}=0.35$ is $F_{sat}=18.6\pm2.5\%$ whereas
\citeauthor{Sawangwit11} measured $F_{sat}=18\pm1$\% for a brighter
selection of LRGs at $z_{earlier}=0.68$. We see that both these
results, for the long-lived model, are significantly higher
compared to the best-fit SDSS HOD, $F_{sat}=8.1\pm1.8$\%. The difference
in the number of the satellite galaxies is explained as the predicted
clustering amplitude at small scales (1-halo term) for the passive
model, is higher compared to the SDSS HOD fit as it is clearly shown in
Fig.~\ref{fig:HODevolve}. Higher clustering signal at small scales
indicates the presence of too many satellite galaxies in the
low-redshift haloes.

The $w(\theta)$ merger model is described by $f_{\rm no-merge}=0.21$ as
presented in Fig.~\ref{fig:HODevolve} and clearly fits the data well.
For this model the satellite fraction at $z=0.35$ estimated to be
$F_{sat}=7.29\pm4.5$\% and is in a good agreement with
\citeauthor{Sawangwit11} Moreover, the best-fit HOD model values for the
evolved $z_{earlier}=1$ LRGs to $z_{later}=0.35$ for bias and galaxy
number density are $b=2.24\pm0.24$ and $n_{g}=0.67\pm0.41 \ 10^{-4} \rm
h^{3}Mpc^{-3}$, respectively. Compared to the SDSS best-fit model, with
$b=2.08\pm0.05$ and $n_{g}=1.3\pm0.4 \ 10^{-4} \rm h^{3}Mpc^{-3}$, the
number of galaxies at $z=0.35$ have been decreased by almost $50\%$ due
to central-central merging. The evolved linear bias and galaxy number
density are consistent with the $z=0.35$ best-fit HOD of
\citeauthor{Sawangwit11} at $1-1.5 \ \sigma$ level.

Note that the agreement at large scales in Fig.~\ref{fig:HODevolve} is
somewhat artificial given the underestimation of $w(\theta)$ by the HOD
model in Fig.~\ref{fig:HOD}b which remains unexplained in the HOD
formalism. But at these smaller scales the result that the merging model
fits better than the long-lived or indeed the virialised clustering
model of Fig.~\ref{fig:wmodel}b may be more robust, given the reasonable
fit of the HOD model at small scales ($\theta < 3'$) in
Fig.~\ref{fig:HOD}b.

\section{Tests For Systematic Errors} \label{sec:tests}

In this section we will present an extended series of checks for
systematic errors that might have affected our clustering
analysis, with the major issue being the flatter slope at large scales as
estimated in \S \ref{sec:plaw}, \S \ref{sec:lcdm_model} and \S \ref{sec:hod}. 
Tests for possible systematics that will be discussed
here are:\\
\begin{itemize}
\item data gradient artefacts, 
\item $w(\theta)$ estimators bias,
\item survey completeness, 
\item observational parameters ; such as star density, galactic extinction, seeing etc.
\end{itemize}

\subsection{Data gradients and $w(\theta)$ estimator bias}
\label{subsec:gradient}

\begin{figure}
 \begin{center} 
\includegraphics[scale=0.35,angle=90]{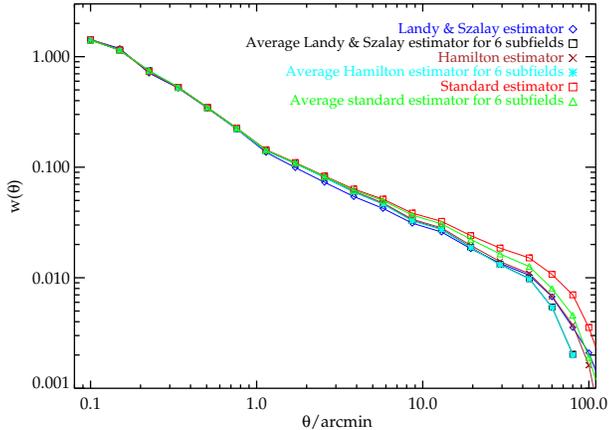} 
\end{center}
 \caption{$w(\theta)$'s from Landy
Szalay, Hamilton and standard estimator of the $700\ \rm deg^{-2}$ LRG sample.
For comparison, the averaged $w(\theta)$'s from the 6 subfields (see text for more detail), 
are overplotted as measured from each estimator.
Landy $\&$ Szalay and Hamilton
estimators, give similar results for the average subfields and the full sample measurements, respectively.
The standard estimator is more biased, at
larger scales.}
\label{fig:all_estim_6sub} 
\end{figure}

A false clustering signal at large scales can arise from artificial gradients 
in the data, as the correlation function is very sensitive
to such factors. In attempting to explain the behaviour of the observed
$w(\theta)$ at large scales, first we divide the LRG sample area in 6
equal subfields in RA. Then the angular correlation function of
each subfield has been calculated using the Landy $\&$ Szalay, Hamilton
and the Peebles estimator - the standard estimator. Furthermore, we
average the $w(\theta)$ results of the 6 subfields as measured by each estimator
and we compare them with $700\ \rm deg^{-2}$ LRG $w(\theta)$ full sample results (see
Fig.~\ref{fig:all_estim_6sub}).

From these comparisons, it is clear that when we use
the Landy $\&$ Szalay and Hamilton estimators, we do not find any
significant difference in the amplitude of the measured $w(\theta)$
between the averaged subfields' or between the full samples' measurements.
When the averaged $w(\theta)$ measurements are compared with those from
the full sample, only a very slightly smaller clustering signal in the averaged 
$w(\theta)$'s is seen, barely visible in Fig.~\ref{fig:all_estim_6sub}. 
Furthermore, this is only the amount expected from the integral constraint (see \S \ref{sec:er_estim})
on $w(\theta)$, if the above Landy $\&$ Szalay estimate is assumed to apply in a single
sub-field area. The standard estimator is known to be subject to larger statistical
errors at large scales and here the signal is actually stronger when
compared with the other two estimators.

\begin{figure}
 \begin{center}
  \includegraphics[scale=0.365,angle=90]{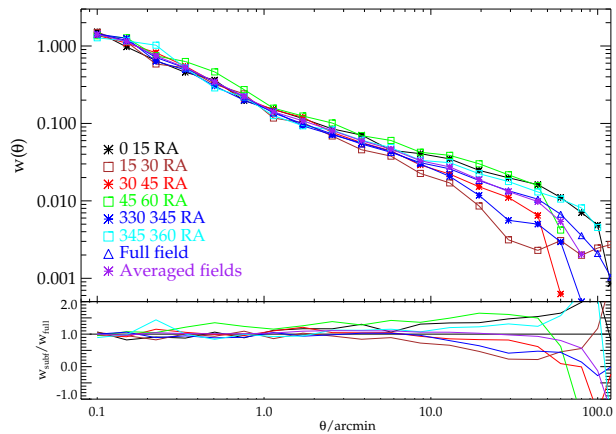} 
\end{center}
\caption{$w(\theta)$ results of the 6 equal size subfields ($15\times 2.5 \rm deg^{2}$ each) across Stripe 82, the total area as estimated by using the Landy Szalay estimator and the averaged clustering 
signal from the 6 subfields. In the bottom panel are displayed the ratios of the $w(\theta)$ of each subfield compared to the total area.}
\label{fig:ls_6subs} 
\end{figure}

Moreover, in Fig.~\ref{fig:ls_6subs} we display the results of the
$w(\theta)$ measurements from the 6 subfields individually against the
full sample measurements as estimated with the Landy $\&$ Szalay
estimator in all cases. Even now we cannot see any major trend through
the subfields' correlation function measurements, except possibly for
the $15^{0} \le RA \le 30^{0}$ subfield which has a steeper slope at larger scales.

\begin{table}
\centering
\begin{tabular}{cc}
\hline\hline
$K$&LRGs $700\ deg^{-2}$ \\
\hline\hline
17.0-17.2  	&    4894 	 	\\
17.2-17.4 		&    11096		\\
17.4-17.6   	&    22490   	\\
17.6-17.8  	&    38659      	 \\
17.8-18.0  	&    53680     	\\
\hline
\end{tabular}
\caption{K-limited sub-samples used for auto-correlations in Fig. 20.}
\label{tab:K_subs}
\end{table}

\begin{figure}
 \begin{center}
 \includegraphics[scale=0.33,angle=90]{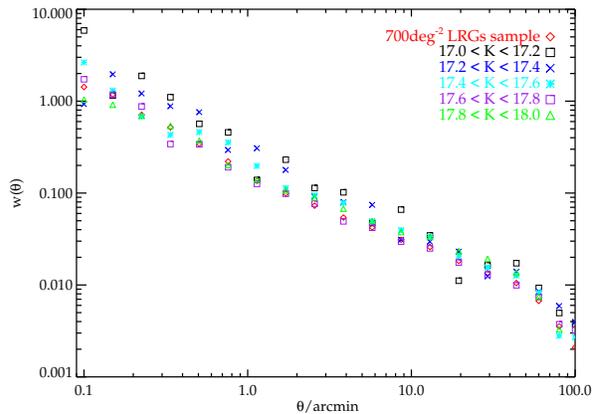}
 \end{center}
 \caption{Auto-correlation functions from Landy-Szalay estimator for the
$700\ deg^{-2}$ LRG K-limited sub-samples from Table 4. Total sample is
overplotted for comparison.}
 \label{fig:K_limit_18}
 \end{figure}

\subsection{Magnitude incompleteness}

Another issue that we want to address is how the clustering signal can
be affected by magnitude incompleteness. The $izK$ colour-selection used for the
LRGs, applied up to the faintest limits of the SDSS-UKIDSS LAS surveys
(see \S \ref{sec:selection}). To account for this, first we divide the
$700\ \rm deg^{-2}$ LRG sample in 5 $K$ magnitude bins in the range $17<K<18$. The number of LRGs in each magnitude
bin is shown in Table~\ref{tab:K_subs}.

   \begin{figure*}
   \begin{center}
   \includegraphics[scale=0.4,angle=0]{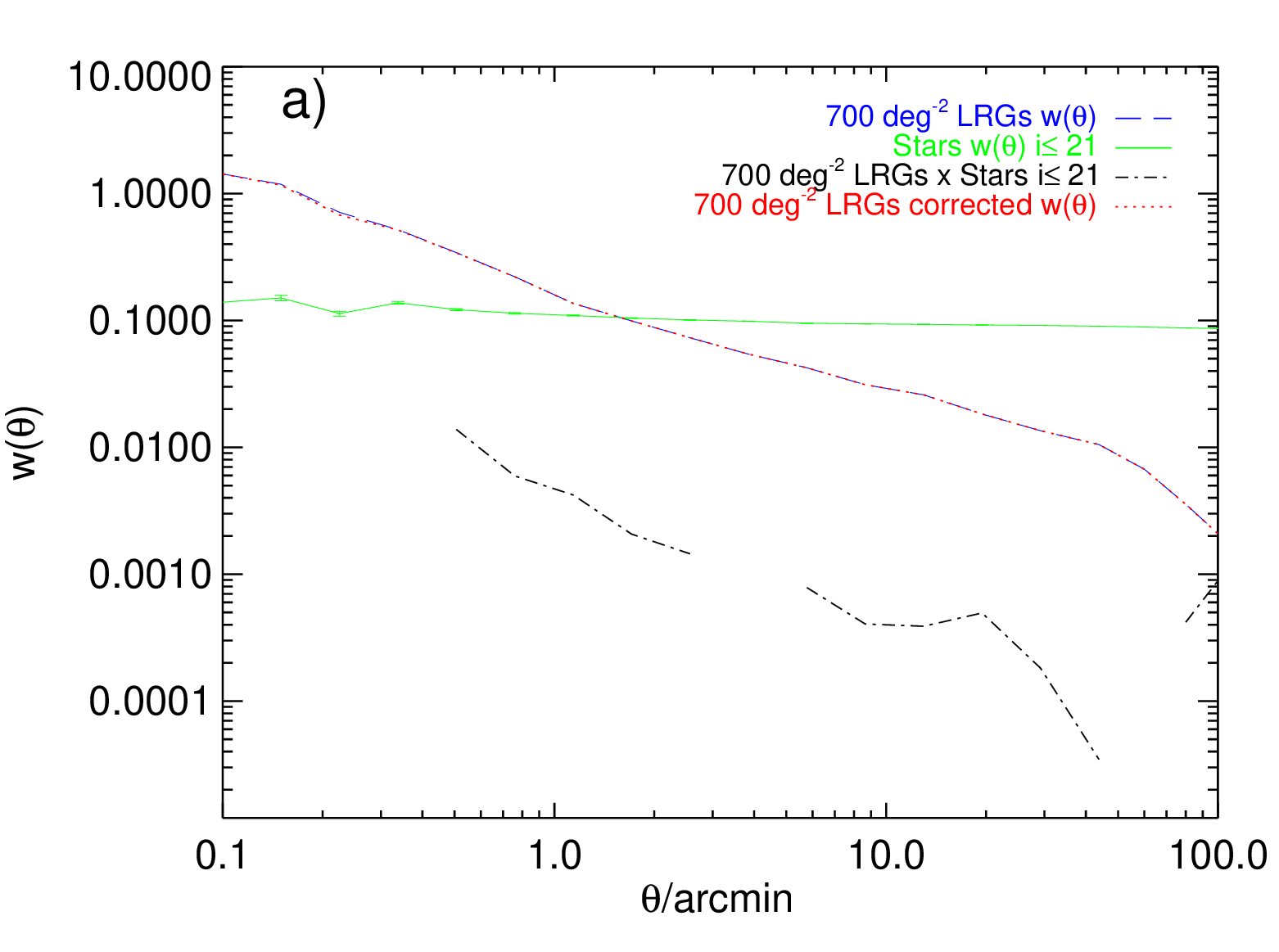}               
   \includegraphics[scale=0.4,angle=0]{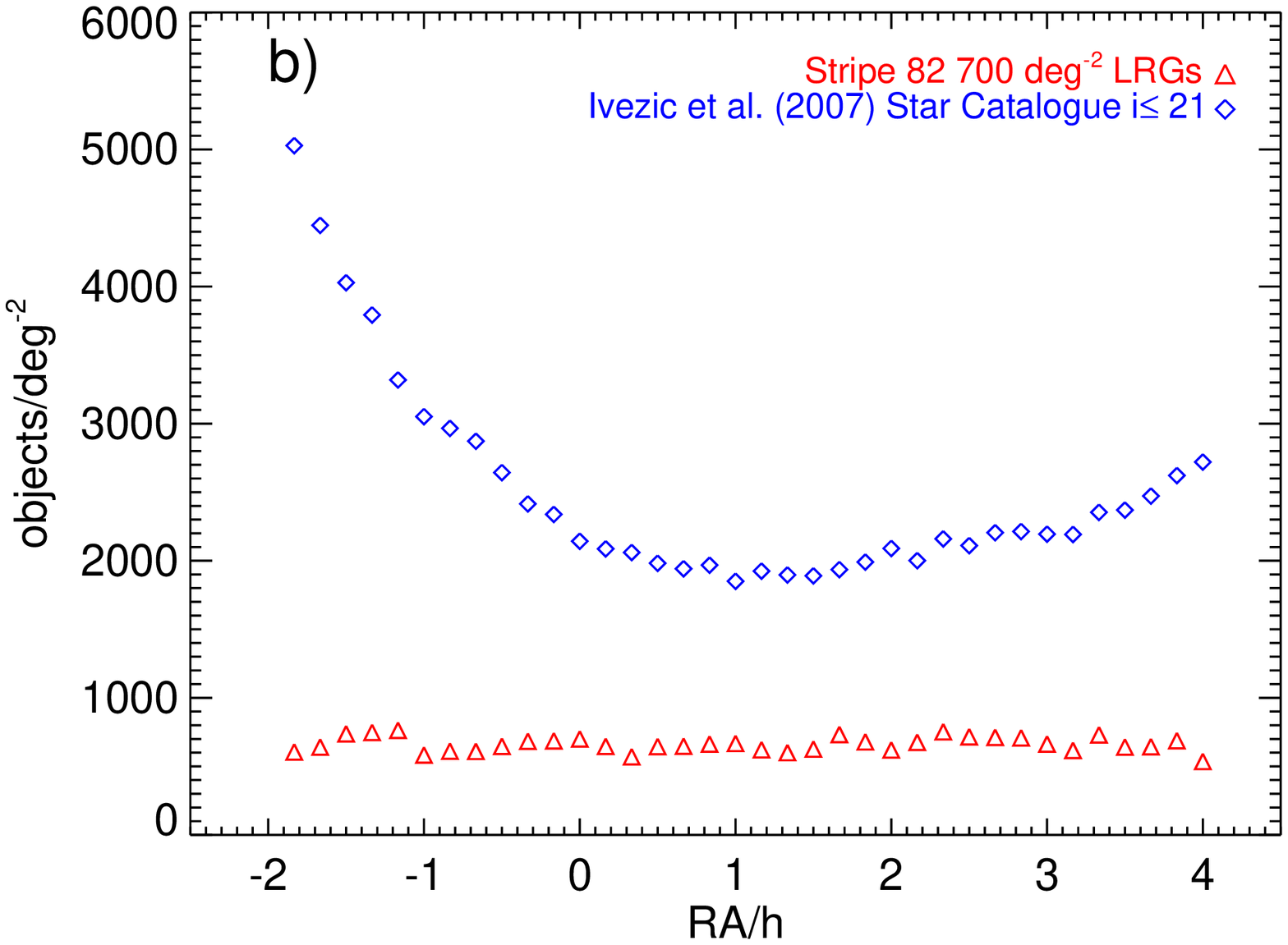} 
   \end{center}
   \caption{\textbf{a)} The observed $w(\theta)$ of Stripe 82 LRGs (blue dashed
   line), Stripe 82 star catalague of Ivezic et al. (2007)
   autocorrelation (green line) for $i\le21$, cross-correlations of the
   aforementioned LRGs-stars (black dashed-dot line) and the resulted
   corrected observed autocorrelation function following Ross et al. (2011). We
   see that there is no difference between the observed LRGs and the
   corrected $w(\theta)$'s, respectively. \textbf{b)} The number density of the
   stars up to $i=21$ from Ivezic et al. (2007) catalogue (blue
   diamonds) and the $700 \rm \deg^{-2}$ LRG sample (red triangles)
   across the Stripe 82. There is a strong gradient in the star
   distribution towards one end of the Stripe 82 at $330\la RA\la340$deg
   or $-2\la RA\la-1$hr in the abscissa notation. But when we excluded this area 
   from the star-LRG cross-correlation, there was no change in the large-scale $w(\theta)$ signal.}
  \label{fig:stars}
  \end{figure*}

  \begin{figure*}
   \begin{center}
   \includegraphics[scale=0.5]{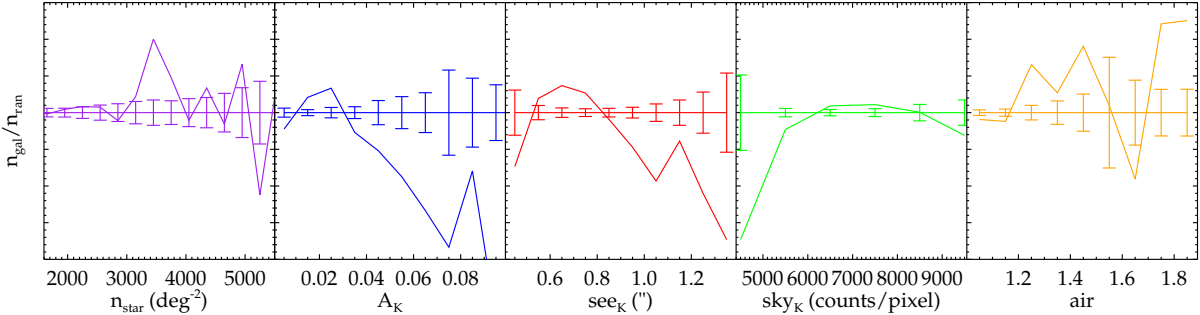}                 
   \end{center}
   \caption{The projected number density of Stripe 82 LRGs as a function of the potential observable systematics: stellar density ($n_{star}$), Galactic 
   extinction ($A_{k}$) in the K-band, the K-band seeing ($see_{K}$), K-band background median sky flux in counts per pixel and the airmass (air). The errors are 
   the standard deviation of the measurements within each bin.}
   \label{fig:stars2}
  \end{figure*}

Measurements of the angular correlation function
from each $K$-bin are shown in Fig.~\ref{fig:K_limit_18}, where
measurement uncertainties are not shown as we are mostly interested in
the shape of the $w(\theta)$ in the linear regime. The clustering
signal from the $K$-magnitude bins compared to the full sample do not
show any significant difference at large scales and follow the full
sample $w(\theta)$ shape. At smaller scales we see that the clustering from the
brighter samples is higher than for the fainter samples, as expected.

The final tests of the magnitude incompleteness check are via the use
of brighter colours in the $zK$ selection. We therefore selected on the
basis of brighter magnitudes down to $z \le 21.2$ and $K \le 17.2$, in
various combinations and re-measured the angular correlation function.
Even with these bright cuts, we did not see any change in the excess at
large scales.

\subsection{Observational parameters}
\label{subsec:obs}
The final test to identify a potential observational systematic effect follows the
approach described by \cite{ARoss11}, referring primarily to the area effectively
masked by stars with magnitudes similar to the galaxies in the field. We
cross-correlated the $700\ deg^{-2}$ LRG sample with the Stripe 82 star
catalogue from \cite{Ivezic07}, in 4 magnitude bins, $i<19.5, 20, 20.5,
21$. From the measured autocorrelation function
of stars and the cross-correlation function of stars with LRGs we computed
the effect of stellar masking on the LRG correlation function using their
equations (28) and (29). We show these results in Fig.~\ref{fig:stars}a.

The cross correlation results show a very small anticorrelation between
LRGs and stars for the $i=19.5$ and $20.5$ bins. A possible explanation for
this anticorrelation might be related to the fact that we see an
increase in the star number density between $330\le RA \le 340$ $deg^{0}$ (see Fig.~\ref{fig:stars}b).
Next, we calculate the expected $w(\theta)$, as defined in Eq. 29 of 
\cite{ARoss11}. In all cases, there was little difference in the expected
and observed $w(\theta)$ of the $700\ \rm deg^{-2}$ LRG sample. We conclude 
that the effect of stellar masking is essentially negligible, less than
1\% of the clustering signal at $\theta\approx90'$

There are other sources of possible systematics as well as star masking.
\cite{ARoss11} also checked observational parameters such as galactic extinction, sky background, seeing and airmass using the cross-correlation technique.
The Stripe 82 LRG sample is K-limited. Hence, we explore if the above observed parameters from the UKIDSS LAS K-band could be sources of systematic errors at large scales.
Fig.~\ref{fig:stars2} displays the number density of Stripe 82 LRGs and how it is related with each potential observational systematic (stars are from \citeauthor{Ivezic07} 2007).
From Fig.~\ref{fig:stars2} we see a sharp decrease in the number of LRGs with high galactic extinction and poor seeing. The airmass fluctuations are also large compared to the error bars. The majority of the LRGs lie within the first few bins of galactic extinction, seeing and airmass in Fig.~\ref{fig:stars2}, but the LRGs in the rest of the bins with higher values could introduce systematics in the clustering signal.

\begin{figure}
 \begin{center}
  \includegraphics[scale=0.35,angle=90]{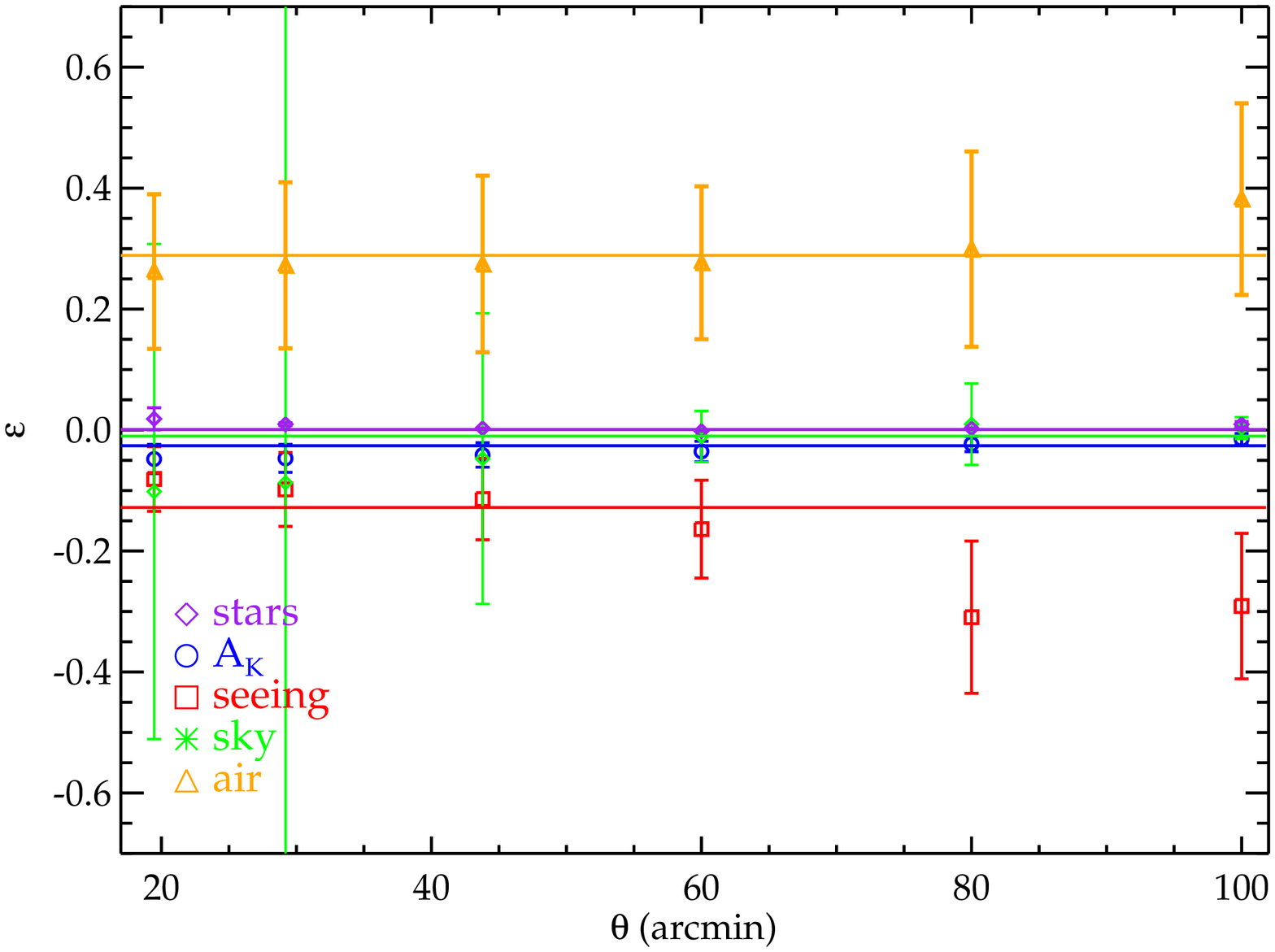} 
  \includegraphics[scale=0.35,angle=90]{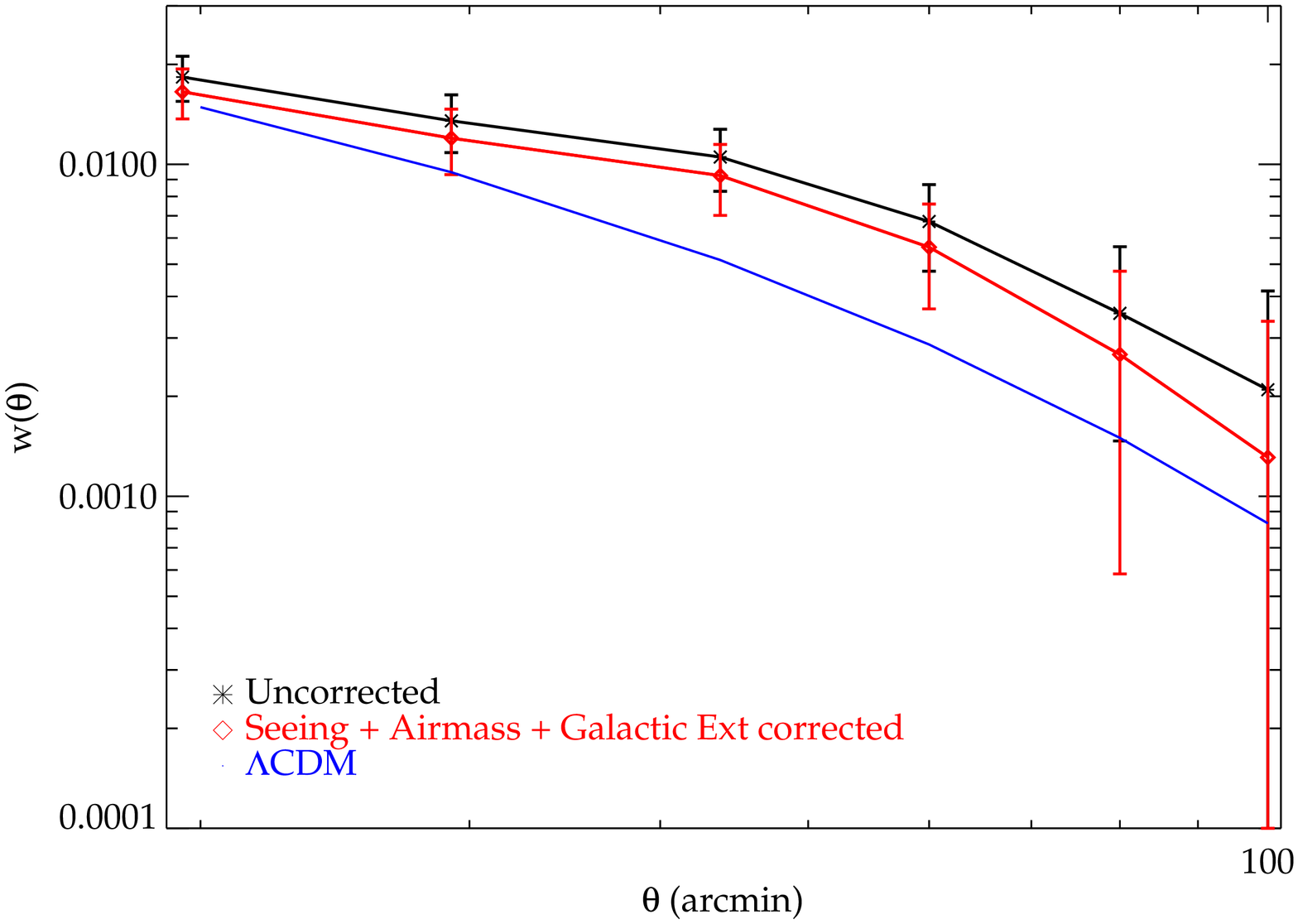} 
  \end{center}
\caption{\textbf{a)} (upper) Similar to Ross et al. 2011 we plot $\epsilon$, the linear factor between the potential observational systematic and its effect on the observed overdensity of galaxies for stars (purple diamond), galactic extinction (blue diamond), seeing (red squares), sky background (green diamond) and airmass (orange triangle). The solid lines are the best-fitting constant value of $\epsilon$ for each systematic. 
   \textbf{b)} (lower) The $w(\theta)$ measurement of the Stripe 82 LRGs without any cross-correlation correction (black star) and $w(\theta)$ corrected for seeing, airmass and galactic extinction combined (red diamond). The best-fit $\Lambda\rm CDM$ model to the uncorrected measurement is plotted (blue line).}
\label{fig:epsilon} 
\end{figure}

\cite{Ho12} present a method to identify which combination of the observed parameters could have the biggest effect on the clustering measurements. 
The authors in this work expressed the linear relationship between the potential observational systematic and its effect on the observed overdensity of galaxies, through the $\epsilon$ factor.
In Fig.~\ref{fig:epsilon}a we show the $\epsilon_{i}$ parameters for each of the the observational parameters.
The \cite{ARoss11} cross-correlation correction technique requires that $\epsilon$ be constant, so we use the best-fitting constant value of $\epsilon$ as calculated with the lowest chi-square fits from field-to-field errors. We find that the biggest correction in the angular correlation function is for the combined seeing, airmass and galactic extinction observational parameters (see Eq. 29 of Ross et al.). Also, a slightly smaller correction has been found for stars, sky background and galactic extinction. In Fig.~\ref{fig:epsilon}b we show the original uncorrected $w(\theta)$ for the Stripe 82 LRGs, the $w(\theta)$ after applying the combined correction for the seeing, airmass and galactic extinction. In the same figure, for comparison we plot the best-fit $\Lambda \rm CDM$ model as displayed in Fig.~\ref{fig:lcdm_model}. 
So far this correction in our $w(\theta)$ results is the most important. But still as we can see from Fig.~\ref{fig:epsilon}b, at the range $20-80'$, the amplitude of the angular correlation function does not show the expected behaviour of the standard model. We have checked for the most common sources of systematics in the literature. Our data could still be affected by hidden artefacts, a case that future studies might be able to identify, but for the moment we will take the corrected result in Fig.~\ref{fig:epsilon}b as our best estimate. Note that the HOD fits of \S \ref{sec:hod} were only done up to 
$\theta \le 60'$ where there is little change in the form of our $w(\theta)$ result.

\section{Test for Non-Gaussianity} 
\label{sec:fnl}
One possible explanation for the flat slope seen at large scales is scale-dependent
bias, although this is usually discussed more in the context of
small-scale clustering. However, scale dependent bias at large scales has previously been
invoked to explain the discrepancy between the APM $w(\theta)$ results and
$\Omega_m=1$ CDM models \citep{Bower93}; in this case the scale dependence
was caused by `cooperative galaxy formation'.

Another possibility is that the LRG power spectrum may be closer to the
primordial power spectrum at higher redshifts. But we have seen that the
Stripe 82 clustering result are not in line with the standard
$\Lambda$CDM model. These correlation function results are better fitted
by a model with $\Omega_m=0.2$ rather than $\Omega_m=0.27$ (see Fig.~\ref{fig:lcdm_model}),
useful at least as an illustration of the size of the LRG clustering excess.

The third possibility is that the $z\approx1$ LRG power spectrum may be
better explained by scale-dependent bias at large scales due to
primordial non-Gaussianity in the density fluctuations. The primordial
non-Gaussianity of the local type is parameterised by $f_{\rm NL}^{\rm
local}$ \cite[see][for a review]{Bartolo04} and is expected to
contribute a $1/k^2$ term to the power spectrum and evolves as $\approx
1+z$ (see Eq.~\ref{eq:nghalobias}). It is therefore best seen at
large-scales and high redshifts. Fig. 1 of \citet{Xia10} shows the potential
effect of non-Gaussianity on the biased clustering of radio sources with
a similar redshift to the LRGs discussed here. It can be seen that the
non-Gaussianity causes a strong positive tail to the correlation
function for $\theta>$ a few degrees.

\citet{Xia10}, following \citet{Blake02}, found that the NVSS
survey ACF showed a strong positive tail suggesting that $f_{NL}^{\rm
local}=62\pm27$. \citet{Xia11} also inspected the angular
correlation function of the DR6 $1\times10^6$ QSO sample and found
similar results to the radio sources with again an extended correlation
function being seen implying similar values of $f_{\rm NL}$ (hereafter
we shall use just $f_{\rm NL}$ to denote $f_{\rm NL}^{\rm local}$) as
for the radio sources. This led to only slightly weaker constraints than
for the radio sources in terms of the value of $f_{\rm NL}$.

\citet{Sawangwit11} measured the combined angular correlation
function of LRGs at $z\approx0.35,0.55,0.68$ and found that although the
results were in agreement with $\Lambda$CDM at scales $<100\rm h^{-1}$Mpc,
at larger scales there was a possible excess, although this could still
be due to systematics.

We then proceeded to follow \citeauthor{Xia10} and fit $f_{\rm NL}$ models. We use
their relation between the non-Gaussian and Gaussian biases ($b_{NG}$
and $b_{G}$) 
\begin{equation}
b_{\rm NG}(z)-b_{\rm G}(z)\simeq2(b_{\rm G}(z)-1)f_{\rm NL}\delta_{\rm ec}(z)\alpha_{\rm M}(k)~. 
\label{eq:nghalobias}
\end{equation}

\noindent Here $\delta_{\rm ec}(z)$ is the critical density for ellipsoidal
collapse and $\alpha_{\rm M}(k)\propto 1/k^2$ contains the scale and
halo mass dependence (see \citeauthor{Xia10} for more details.)

   \begin{figure}
 \begin{center}
 \includegraphics[scale=0.45]{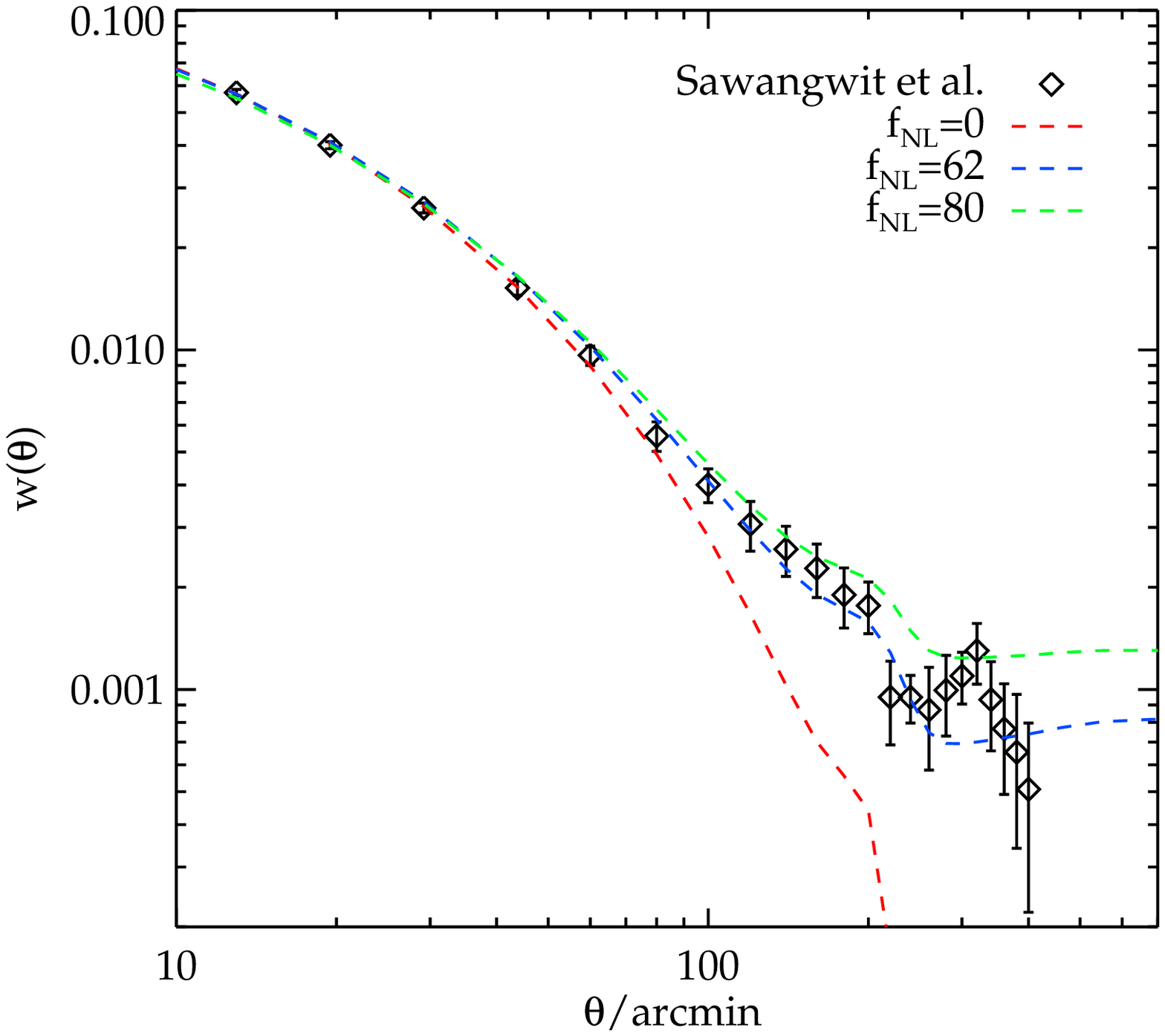}
 \includegraphics[scale=0.35,angle=90]{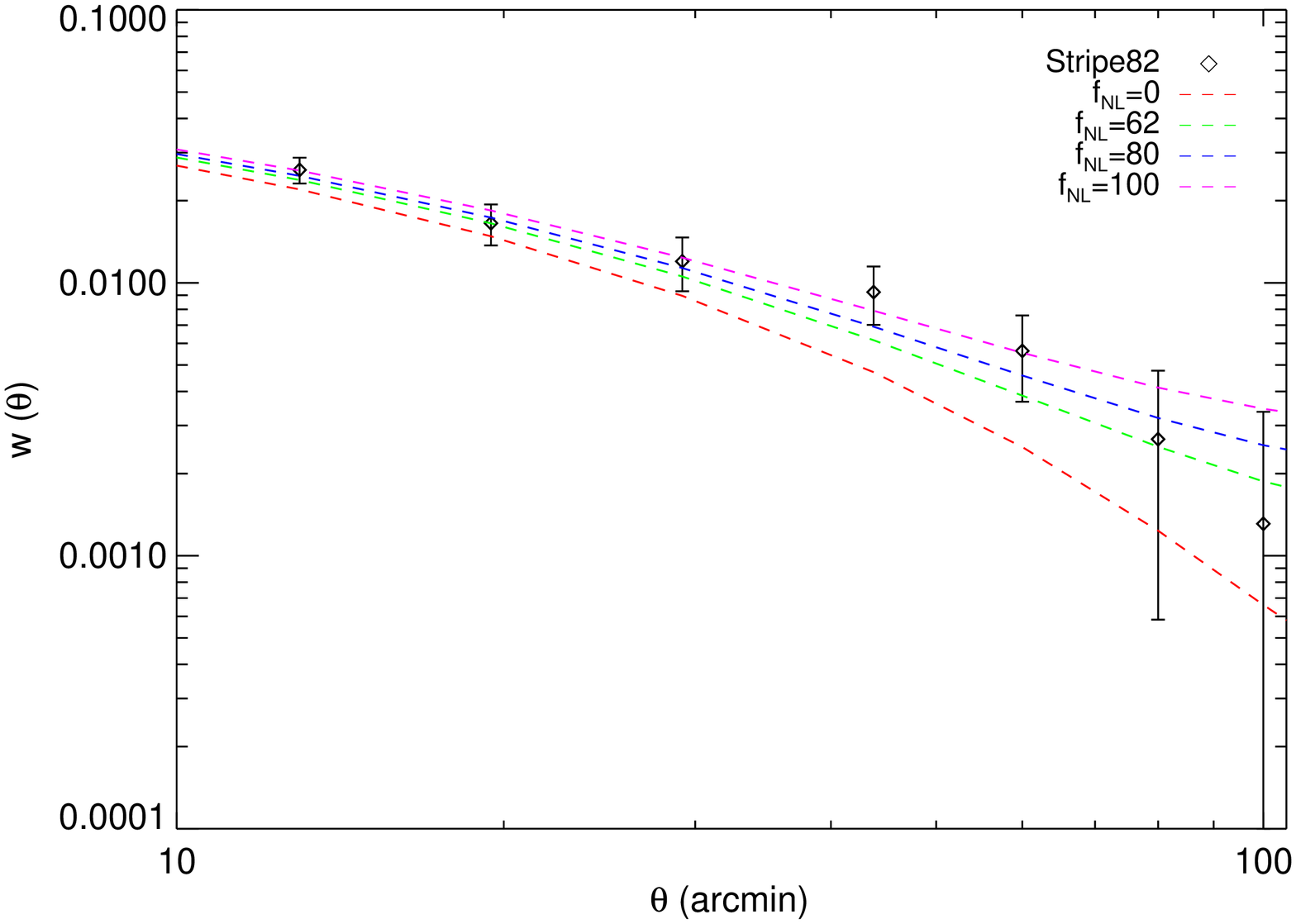}
 \end{center}
 \caption{\textbf{a)} (upper) The combined correlation function of Sawangwit et al (2011) for the
$z=0.35$, $z=0.55$ and $z=0.68$ LRG  samples, compared to a standard LCDM
model ($f_{nl}=0$) and models with increasing degrees of primordial
non-Gaussianity ($f_{nl}=62, 80$). \textbf{b)} (lower) The Stripe-82 $z\approx1$ LRG correlation function compared to a
standard LCDM model ($f_{nl}=0$) and models with increasing degrees of
primordial non-Gaussianity ($f_{nl}=62, 80, 100$).}
 \label{fig:LRG_fnl}
 \end{figure}

We first applied this relation to the case of the NVSS radio sources at $z\approx0.7$. We
found that adding the $1/k^2$ term to the standard cosmology $P(k)$
caused it to diverge and so we had to apply a large-scale cut-off, so
that for $k<k_0$ then $P(k)=0$. This is clearly a source of uncertainty
in fitting for $f_{\rm NL}$. Nevertheless, we found that for $k_0=10^{-6}$, we
could reproduce the results of \citet{Xia10}.

We then applied the same technique and cut-off to the combined $AA\Omega$
LRG and the Stripe 82 LRG $w(\theta)$'s (after applying the combined correction for seeing, airmass and galactic extinction as estimated in \S~\ref{subsec:obs}). We first took the value of $b_G=2.08$
from the halo model fits of \citet{Sawangwit11} and fitted for $f_{NL}$.
The result is shown in Fig.~\ref{fig:LRG_fnl}a. We find that for
$AA\Omega$ LRGs, the results for $f_{NL}$ are reasonably compatible with
those from the NVSS catalogue with values of $f_{\rm NL}=60-80$ giving a
better fit to the data in the range $1.5<\theta<6.5$deg.

The prediction from non-Gaussianity is that the large scale slope will
further flatten with redshift. We therefore compared the Stripe 82 LRGs
to models with the same $f_{\rm NL}$ values and find no inconsistency
(see Fig.~\ref{fig:LRG_fnl}b). Clearly the errors at the largest scales
are more significant for the Stripe 82 data than for the $AA\Omega$ LRGs
or the NVSS radio sources. However, the predicted flattening of the
Stripe 82 correlation function at $\theta\approx1$deg makes the
non-Gaussian models more consistent with the data in this smaller
angular range than the $f_{\rm NL}=0$ model. At larger scales the errors
are larger and the data is therefore more in agreement with the standard
model.

  \begin{figure}
 \begin{center}
 \includegraphics[scale=0.45]{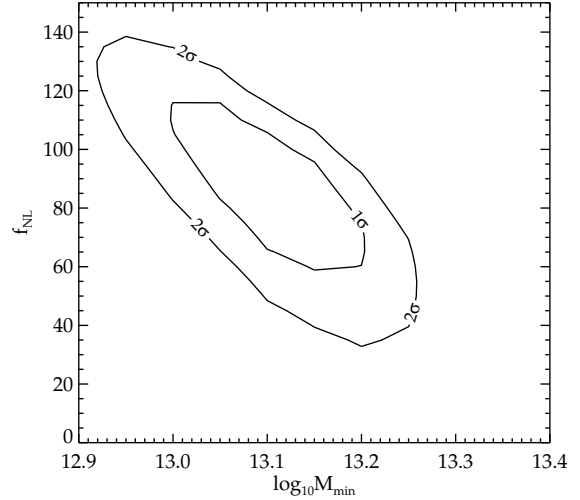}
 \end{center}
 \caption{The minimum $\chi^2$ is 5.5 over 11 d.o.f and
the best-fit parameters are $f_{\rm NL}=90\pm30$ $(1\sigma)$ and
$M_{min}=1.26\pm0.22\times10^{13} \rm h^{-1} M_{\odot}$. The best-fit
$M_{min}$ here is lower than the full HOD fit assuming $f_{NL}=0$ at 
$2.2 \times 10^{13}\rm
h^{-1}M_{\odot}$. }
 \label{fig:fNL}
 \end{figure}

Fig. ~\ref{fig:fNL} shows the effect of jointly fitting $f_{NL}$ on the
minimum halo mass, $M_{min}$, in the HOD model. The best fit
model now gives $M_{min}=(1.26\pm0.22)\times10^{13} h^{-1} M_{\odot}$ and
$f_{NL}=90\pm30$, lower than then the $M_{min}=2.2 \times
10^{13}h^{-1}M_{\odot}$ value when $f_{nl}=0$ is assumed in the full HOD
fit.

We should say that rather than detections of non-Gaussianity, the
present $AA\Omega$ and Stripe 82 LRG results should be more regarded as upper
limits on non-Gaussianity. Large-scale angular correlation function
results are still susceptible to large-scale gradients and even though
there is no direct evidence for these in the $AA\Omega$ or Stripe 82 samples,
there is still the possibility that these exist in the data. On the
other hand, the classic test for the reality of a correlation function
feature is that it scales correctly with depth and at least the SDSS and Stripe 82 LRG correlation functions in 
Figs.~\ref{fig:LRG_fnl}a,b look like they do so. It will be
interesting to see if as QSO surveys \citep{Sawangwit12} and $z\approx3$ LBG
surveys \citep{Bielby12} grow, whether the correlation
functions at higher redshift also show an increased slope flattening as
predicted for the non-Gaussian models.

The other uncertainty that has arisen is in the non-Gaussian model
itself where we have found that there is a rather strong dependence on a
small-scale cut-off, $k_0$. Other authors have made some reference to
this problem but only implicitly. It will be interesting to see if more
accurate models for non-Gaussianity can numerically predict this cut-off
from first principles.

\section{Summary and conclusions}

We have measured $w(\theta)$ for $\approx130\,000$ colour selected galaxies in
Stripe 82 exploiting SDSS DR7 $i+z$ bands and UKIDSS LAS $K$ photometry. We used
the cross-correlation technique of Newman (2008) to establish that the
average redshift of the LRGs is $z\approx1$. This sample therefore
probes higher redshifts than the previous SDSS LRG samples of
\citet{Sawangwit11}. We have established that a sample with sky density
$\approx700$deg$^{-2}$ has a comparable space density to the
$z\approx0.68$ $AA\Omega$ LRG sample of \citet{Sawangwit11}. However, this
is only an approximate correspondence which makes evolutionary
comparisons between the redshifts more tricky. What is clear is that the
$z\approx1$ LRGs generally have a relatively high clustering amplitude. Compared
to the $AA\Omega$ LRG $w(\theta)$ scaled to the depth of the Stripe 82 LRGs, the Stripe 82
$w(\theta)$ is higher at all scales, even those below $<1\ \rm h^{-1}$Mpc. Thus at
intermediate scales, the $z\approx1$ LRGs are not only more clustered
than predicted by the long-lived evolutionary model, they are also more
clustered than the comoving model. At small separations ($\la 1\ \rm h^{-1}$Mpc) the
correlation function amplitude is again somewhat higher than the $AA\Omega$
results scaled by the previously preferred stable clustering model. The
Stripe 82 $w(\theta)$ also shows a very flat slope at large scales which means that
the $\Lambda \rm$CDM linear model has become a poorer fit than at lower redshift.

Partly to look for an explanation for the flat large-scale slope, we
then fitted a HOD model to the Stripe 82 $w(\theta)$. The best fit parameters were
$M_{min}=2.19\pm0.63\times10^{13}\ \rm h^{-1}M_{\odot}$, $M_1=21.9\pm5.6\times10^{13}\ \rm h^{-1}M_{\odot}$,
$b_{\rm lin}=2.81\pm0.18$, $M_{\rm eff}=3.3\pm0.6\times10^{13}\ \rm h^{-1}M_{\odot}$, $F_{\rm sat}=3.17\pm0.08\%$ and $n_{\rm g}=0.8\pm0.3\times10^{-4}\ \rm h^{3}Mpc^{-3}$.
The high amplitude of the correlation function
clearly pushes the halo masses up and the space densities down. The
lowest chi-square fits were found when large scales were excluded but
the reduced chi-squares were still in the range 2.3-3.6. This is
actually an improvement over the lower redshift samples but this is
certainly due to the larger errors on the Stripe 82 data. We conclude that it
is not possible to find an explanation for the flat slope in the Stripe 82
$w(\theta)$ on the basis of the HOD model.

We also then studied the evolution of the HOD between $z=1$ and $z=0.35$.
Similar to \citet{Sawangwit11}, we concluded that a pure passive model
with a low merger rate might produce too steep a $w(\theta)$ slope at small
scales ($<1\ \rm h^{-1}Mpc$). In this case, we have already noted that the
small scale amplitude may also be too high for a passive model with
stable clustering.

We have looked for an explanation of the flat slope in terms of systematics by cross correlating 
the Stripe 82 LRG sample with stellar density, airmass, seeing, sky background and galactic extinction and used the method of \citet{ARoss11} to correct our $w(\theta)$. Even the combined correction
for seeing, airmass and galactic extinction only produced a small change in $w(\theta)$ at large scales.

We conclude that the high amplitude and flat slope of the Stripe 82 LRGs
$w(\theta)$ may have a significant contributions from the uncertainty in
the comparison between $AA\Omega$ and Stripe 82 LRG luminosities. However, this
leaves a similar contribution from a new and unknown source. We have
discussed large-scale, primordial, non-Gaussianity as one possibility for the 
source of this large-scale excess. We have suggested that the evidence
from the $AA\Omega$ sample itself for an excess at even larger scales may
fit in with the behaviour expected from non-Gaussianity over this
redshift range. In this case we returned to the fitting of halo masses including the
non-Gaussian component and found that the best fit $M_{min}$ decreased
from $2.2\times10^{13}$M$_\odot$ to $1.3\times10^{13}$M$_\odot$. More
importantly, if the Stripe 82 large-scale $w(\theta)$ excess proves
reliable and not due to systematics, then we have made a significant
detection of non-Gaussianity in the $z\approx1$ LRG distribution with an
estimated local non-Gaussianity parameter estimate of $f_{\rm NL}^{\rm
local}=90\pm30$. This represents a $3\sigma$ detection at a level
comparable to the present upper limit from WMAP CMB measurements of
$f_{\rm NL}^{\rm local}=32\pm21$ \citep{Komatsu10}.

\section*{Acknowledgements}
NN acknowledges receipt of a fellowship funded by the European Commission's Framework Programme 6, through the Marie Curie Early Stage Training project MEST-CT-2005-021074.
US acknowledges financial support from the Institute for the 
Promotion of Teaching Science and Technology (IPST) of The Royal 
Thai Government.

We would like to thank Dr. Nigel Hambly and the wsa-support team for their help with UKIDSS data.
The UKIDSS project is defined in Lawrence et al 2007. UKIDSS uses the 
UKIRT Wide Field Camera (WFCAM; Casali et al 2007) and a photometric 
system described in Hewett et al 2006. The pipeline processing and science 
archive are described in Irwin et al (2008) and Hambly et al (2008).

Funding for the SDSS and SDSS-II has been provided by the Alfred P.
Sloan Foundation, the Participating Institutions, the National Science
Foundation, the U.S. Department of Energy, the National Aeronautics and
Space Administration, the Japanese Monbukagakusho, the Max Planck
Society, and the Higher Education Funding Council for England. The SDSS
Web Site is http://www.sdss.org/.

The SDSS is managed by the Astrophysical Research Consortium for the
Participating Institutions. The Participating Institutions are the
American Museum of Natural History, Astrophysical Institute Potsdam,
University of Basel, University of Cambridge, Case Western Reserve
University, University of Chicago, Drexel University, Fermilab, the
Institute for Advanced Study, the Japan Participation Group, Johns
Hopkins University, the Joint Institute for Nuclear Astrophysics, the
Kavli Institute for Particle Astrophysics and Cosmology, the Korean
Scientist Group, the Chinese Academy of Sciences (LAMOST), Los Alamos
National Laboratory, the Max-Planck-Institute for Astronomy (MPIA), the
Max-Planck-Institute for Astrophysics (MPA), New Mexico State
University, Ohio State University, University of Pittsburgh, University
of Portsmouth, Princeton University, the United States Naval
Observatory, and the University of Washington.

Funding for the DEEP2 survey has been provided by NSF grants
AST95-09298, AST-0071048, AST-0071198, AST-0507428, and AST-0507483 as
well as NASA LTSA grant NNG04GC89G.

 \bibliographystyle{mn2e}

\end{document}